\documentclass[aip,cha, reprint,showpacs,showkeys]{revtex4-1}
\usepackage{amssymb, dcolumn,  graphicx, latexsym, amsfonts, bm}

\begin{document}

\title{Interaction of the fast wave with a scrape-off layer plasma in tokamaks.\\ 
The ion cyclotron parametric instabilities and the anomalous heating of ions}

\author{V. S. Mikhailenko}\email[E-mail:]{vsmikhailenko@pusan.ac.kr }
\affiliation{Plasma Research Center, Pusan National University,  Busan 46241, South Korea}
\author{V. V. Mikhailenko}\email[E-mail: ]{vladimir@pusan.ac.kr}
\affiliation{Plasma Research Center, Pusan National University,  Busan 46241, South Korea}
\author{Hae June Lee}\email[E-mail: ]{haejune@pusan.ac.kr}%
\affiliation{Department of Electrical Engineering, Pusan National University, Busan 46241, 
South Korea}
\date{\today}

\begin{abstract}
The theory of the ion cyclotron (IC) electrostatic parametric instabilities of plasma which 
are driven by the elliptically 
polarized fast wave (FW) of the finite wavelength is developed. This theory grounds on the 
methodology of the oscillating modes. It provides the analytical treatment of the 
parametric instabilities with wavelength comparable with the displacements of ions relative 
to electrons in FW. The growth rate of the IC quasimode decay instability, which 
was considered as a potential source of the generation of the high energy ions in scrape-off layer (SOL) during FW injection in tokamaks, is derived analytically for arbitrary 
values of the FW electric field and wavelengths of the unstable IC perturbations. 
The comprehensive numerical analysis of the dispersion equation for three wave system which contains the IC mode $\varphi_{i}\left(\mathbf{k}_{i}, \omega\right)$ and its harmonics $\varphi_{i}\left(\mathbf{k}_{i},  \omega-\omega_{0}\right)$, $\varphi_{i}
\left(\mathbf{k}_{i}, \omega-2\omega_{0}\right)$, where $\omega_{0}$ is FW frequency, is performed. It reveals that the parametric  IC instability for this wave system has the maximum growth rate for the IC waves with wavelength  comparable with the thermal ion Larmor radius. It is found that the inverse electron Landau damping plays essential role in the development of this instability. 
The possible mechanism of the saturation of this instability is the scattering of ions by the 
ensemble of the IC waves with random phases, which limits the development of the instability 
on the high level. The anomalous heating rates of ions resulted from the interactions of ions with parametric IC turbulence is determined employing the developed  quasilinear theory for the IC quasimode decay instability. The derived results reveals, that the experimentally 
observed anisotropic heating of cold SOL ions may be caused by the parametric IC 
turbulence in SOL. However, the IC parametric turbulence is unlikely to be responsible for the
experimentally observed bursts of poorly confined suprathermal ions in the SOL of tokamak plasmas. 
\end{abstract}
\pacs{52.35.Ra, 52.35.Kt}

\maketitle

\section{Introduction}\label{sec1}
The understanding of interaction between the fast  wave (FW) in the ion cyclotron range of frequency (ICRF) and the cold 
scrape-off layer (SOL) plasma  adjacent to the ICRF antenna is crucial in FW heating experiments. It was found in essentially all these experiments 
that up to 50$\%$ or more\cite{Perkins1,Perkins2,Bertelli} of the 
launched FW power is lost in the SOL outside of the last closed flux surface. This significant fractions of the coupled 
FW power is deposited eventually to the inner wall and divertor plates of tokamaks. 

The bursts of poorly confined energetic ions with energy above 20 keV, which is observed regularly immediately following FW injection on 
DIII-D tokamak, is one of the well 
recognized channel of the FW energy leakage\cite{Pace}. It was experimentally proved in Ref. \cite{Pace} that these high 
energy ions in DIII--D tokamak were created  within the SOL. The other channel of the FW 
energy absorption, which also is absent within linear treatment, is a significant 
heating of the low energy ions in SOL \cite{Wilson}, which was found to be anisotropic with 
$T_{i\bot}\gg T_{i\parallel}$. It was claimed in Refs. \cite{Pace, Wilson} that the observed 
anomalous absorption of the FW energy by the SOL ions is the result of their interactions 
with the IC parametric quasi-mode decay instabilities\cite{Porkolab1, Porkolab2} which were 
detected in the SOL near the antenna strap. 

The excitation of the parametric instabilities in SOL during the ICRF heating of plasma when a RF power threshold was exceeded 
is observed regularly  and is well investigated experimentally\cite{Pace,Wilson,Porkolab1, Porkolab2,Nieuwenhove,Fujii,Rost}. However, the clear understanding the role of the parametric instabilities in the FW
power losses in SOL is still missing and there is still lack a verification by the theory.


In this paper, we consider the theory of the ion cyclotron (IC) parametric instabilities 
driven by FW and the resulted anomalous heating of ions in 
the SOL plasma. The special attention will be given to the linear and quasilinear theory of 
the quasimode decay instabilities development of which 
is considered \cite{Pace,Wilson,Fujii,Rost} as the main mechanism of the 
transmission of the FW energy to the SOL plasma and particularly to the SOL ions.
The investigations of the IC parametric turbulence in the present paper bases on the 
methodology of the oscillating modes, developed in Refs.\cite{Mikhailenko1,Mikhailenko4}, 
and extended in this paper on the accounting for the finite 
wavelength of FW. In Sec. \ref{sec2}, we present in details the justification of the local 
approximation in the theory of the parametric instabilities powered by the FW of the finite 
wavelength which we use in our theory instead of the usually applied approximation of the  
spatially homogeneous FW (dipole approximation). The 
basic equations which govern the electrostatic parametric instabilities driven by FW is 
developed in Sec. \ref{sec3}. In the Subsec. \ref{sec3.1} of this section, the basic 
dispersion equation for the electrostatic parametric 
instabilities driven by FW is derived. The short summary of the theory of the IC kinetic 
parametric instability, which is the simplest application of the derived dispersion equation 
is presented in Subsec. \ref{sec3.2}. The linear theory of the quasimode decay instability 
\cite{Porkolab1, Porkolab2} is presented in Subsec. \ref{sec3.3}. The main purpose of our 
analysis of the quasimode decay instability is to elucidate the 
role of these instabilities in the turbulent heating of ions and generation of the population 
of the high energy ions in the SOL during FW heating.
This problem is considered in Sec. \ref{sec4}, where the rates of the anomalous heating of 
ions resulted from their interaction with IC 
turbulence powered by the IC quasimode decay instability are determined. Conclusions are 
given in Sec. \ref{sec5}.

\section{The local approximation in the theory of FW interaction with plasma}\label{sec2}

In this section, we derive the basic equations which govern the interaction of the high power 
FW with a near-antenna scrape-off layer (SOL) plasma, where the oscillatory velocity of ions 
in FW field is commensurate with the ion thermal velocity. 
We use a slab geometry with the mapping $\left(r,\theta, \varphi\right)\rightarrow 
\left(\hat{x}, \hat{y}, \hat{z}\right)$ where $r$, $\theta$,
$\varphi$ are the radial, poloidal and toroidal directions, respectively, of the toroidal 
coordinate system. The components of the electric and magnetic 
fields, $\mathbf{E}_{0}\left(\mathbf{\hat{r}}, t\right)$, $\mathbf{\tilde{B}}_{0}
\left(\mathbf{\hat{r}}, t\right)\sim e^{ik_{0x}\hat{x}+ik_{0y}\hat{y}
+ik_{0z}\hat{z}-i\omega_{0}t}$ of the elliptically polarized fast  wave are such that $E_{0x}
\sim E_{0y}\gg E_{0z}$, 
$k_{0x}\sim k_{0y} \gg k_{0z}$, $\tilde{B}_{0x}\sim \tilde{B}_{0y} \ll \tilde{B}_{0z}$ and 
component $\tilde{B}_{0z}$ is much less than the magnitude of the confined magnetic field $
\mathbf{B}$ 
pointed along coordinate $\hat{z}$. Therefore we consider a model of a plasma immersed in the 
uniform magnetic field $\mathbf{B}$ directed along 
coordinate $\hat{z}$ and in the fast wave electric field of the form
\begin{eqnarray}
&\displaystyle
\mathbf{E}_{0}\left(\mathbf{\hat{r}}, t\right)=\mathbf{E}_{0x}\cos\left(\omega_{0}t-
\mathbf{k}_{0}\mathbf{\hat{r}}\right)+\mathbf{E}_{0y}\sin
\left(\omega_{0}t-\mathbf{k}_{0}\mathbf{\hat{r}}\right)
\label{1}
\end{eqnarray}
with wave vector $\mathbf{k}_{0}=\mathbf{k}_{\bot}+\mathbf{k}_{0z}$ where $\mathbf{k}_{\bot}=
\mathbf{k}_{0x}+\mathbf{k}_{0y}$ 
is directed perpendicular to the magnetic field $\mathbf{B}$, and $\mathbf{k}_{0z}$ is 
pointed along $\mathbf{B}$. Our theory bases on 
the Vlasov equation for the velocity distribution function $F_{\alpha}$ of $\alpha$ plasma 
species ($\alpha = i$ for ions and $\alpha = e$ for electrons),
\begin{eqnarray}
&\displaystyle \frac{\partial F_{\alpha}}{\partial
t}+\left. \mathbf{\hat{v}}\frac{\partial F_{\alpha}}
{\partial\mathbf{\hat{r}}}+\frac{e}{m_{\alpha}}\right(\mathbf{E}_{0}\left(\mathbf{\hat{r}}, t\right)
\nonumber 
\\ 
&\displaystyle
\left.+\frac{1}{c}\left[\mathbf{\hat{v}}\times
\mathbf{B}\right]-\bigtriangledown
\varphi\left(\mathbf{\hat{r}},t\right)\right)\frac{\partial
F_{\alpha}}{\partial\mathbf{\hat{v}}}=0,\label{2}
\end{eqnarray}
and on the Poisson equation for the self-consistent electrostatic potential $\varphi
\left(\mathbf{\hat{r}}, t\right)$ of the plasma respond on the fast wave,
\begin{eqnarray}
&\displaystyle 
\vartriangle \varphi\left(\mathbf{\hat{r}},t\right)=
-4\pi\sum_{\alpha=i,e} e_{\alpha}\int f_{\alpha}\left(\mathbf{\hat{v}},
\mathbf{\hat{r}}, t \right)d{\bf \hat{v}}. \label{3}
\end{eqnarray}
In Eq. (\ref{3}), $f_{\alpha}$ is the fluctuating part of the distribution function 
$F_{\alpha}$, $f_{\alpha}=F_{\alpha}-F_{0\alpha}$, where $F_{0\alpha}$ is the equilibrium distribution function. 

The theory of the electrostatic plasma instabilities and plasma turbulence is grounded, as a rule, on the application 
to the governing equations of the spectral (Fourier, Laplace) transforms over time and spatial coordinates, 
which follows by the investigation of the spectral properties, stability and temporal evolution of the separate 
spatial Fourier mode of the 
perturbed electrostatic potential. The principal obstacle for the direct application of the spectral transforms to Eq. 
(\ref{1}) is the explicit time and coordinate dependencies of the spatially inhomogeneous electric field of the 
applied electromagnetic wave. It is commonly accepted in the theoretical investigations of the parametric 
instabilities excited by the strong driving FW, that the 
approximation of the spatially homogeneous pump wave with $\mathbf{k}_{0}=0$ across the magnetic field may suffice 
since the parametrically excited waves 
have the wave number across the magnetic field much larger than the wave number of the pump wave. Direct application 
of the spectral transform over time 
variable to Vlasov-Poisson system of equations in this case results in the set of the coupled equations for the 
infinite number of the harmonics $f_{\alpha}\left(\mathbf{k}, \omega-n\omega_{0}
\right)$ where $n=0, \pm 1, \pm 2,..$, coupled with infinite numbers of harmonics 
of the perturbed electrostatic potential $\varphi\left(\mathbf{k}, \omega-m\omega_{0}\right)$ 
where $m=0, \pm 1, \pm 2,..$. Usually\cite{Silin,Porkolab}, in order to simplify the problem, the transformation to the coordinate 
frame in velocity space
which oscillate with velocity $\mathbf{V}_{i}\left(t\right)$ of ions (electrons), but with unchanged coordinates in configuration 
space, was employed; specifically, the transformation is given by $t'=t$, $\mathbf{r}'=\hat{\mathbf{r}}$, $\mathbf{v}=\hat{\mathbf{v}}
-\mathbf{V_{0}}\left( \hat{\mathbf{r}}\right)$. In new coordinates, the spatially uniform RF electric field is excluded from the Vlasov
equation for ions (electrons). Then, the application of the spectral transform over these 
time and spatial coordinates to this Vlasov equation gives the known equation for the 
separate spatial Fourier mode of the perturbations of the ion (electron) 
distribution functions  coupled with the separate spatial Fourier mode of the perturbed 
electrostatic potential with the same wave number and frequency. The Fourier transform of the Poisson equation for the perturbed electrostatic potential 
is performed in this approach in the laboratory frame. In the laboratory frame, the 
perturbations of the ion density determined in the ion oscillation frame and of the electron density, 
determined in the electron oscillating frames, are observed in the laboratory frame as the  infinite number of the harmonics 
$\varphi\left(\mathbf{k}, \omega-m\omega_{0}\right)$ 
where $m=0, \pm 1, \pm 2...$, of the perturbed potential. Because of this, the analytical investigations 
of the dispersion properties of the separate Fourier mode of the perturbed potential becomes 
possible only for the limiting  case of the small displacements 
of the particles with respect to the wave length of the perturbations\cite{Porkolab}. It was derived \cite{Mikhailenko1} that  
the Poisson equation Fourier transformed in the ion or electron oscillating frame, instead of the laboratory frame, admits 
the investigations of the  stability of the perturbed potential for the displacements of 
particles in the RF field comparable with the wavelength of the unstable waves.

In this paper, we extend the approach developed in Ref. \cite{Mikhailenko1} on the RF wave with finite wave vector $\mathbf{k}_{0}$ assuming that the 
displacements $\mathbf{\xi}_{i,e}$ of the ions/electrons in the RF field is much less than the fast wave wavelength, 
i.e. that $\mathbf{k}_{0}\cdot\mathbf{\xi}_{i,e}\ll 1$. We transform the Vlasov equation (\ref{2}) for ions to new position 
vector $\mathbf{r}_{i}$ and velocity $\mathbf{v}_{i}$, determined by the relations
\begin{eqnarray} 
&\displaystyle
\hat{t}=t,\qquad \hat{\mathbf{v}}=\mathbf{v}_{i}+\mathbf{V}_{i}
\left(\mathbf{r}_{i}, t\right), 
\nonumber 
\\ &\displaystyle
\hat{\mathbf{r}}=\mathbf{r}_{i}+\mathbf{R}_{i}\left(\mathbf{r}_{i}, t\right)=\mathbf{r}_{i}+\int\limits^{t}
_{t_{(0)}}\mathbf{V}_{i}
\left(\mathbf{r}_{i}, t_{1}\right)dt_{1},\label{4}
\end{eqnarray}
or
\begin{eqnarray}
&\displaystyle
\mathbf{v}_{i}=\hat{\mathbf{v}}-\hat{\mathbf{V}}_{i}\left(\hat{\mathbf{r}},
\hat{t}\right), \,\,\,\mathbf{r}_{i}=\hat{\mathbf{r}}-\int\limits^{\hat{t}}_{t_{(0)}}\hat{\mathbf{V}}_{i}
\left(\hat{\mathbf{r}}, \hat{t}_{1}\right)d\hat{t}_{1},\label{5}
\end{eqnarray}
determined by the frame of reference moving with velocity
$\hat{\mathbf{V}}_{i}\left(\hat{\mathbf{r}}, \hat{t} \right) =
\mathbf{V}_{i}\left(\textbf{r}_{i}, t \right)$ relative to the laboratory frame. 
In new variables, the Vlasov equation has a form
\begin{eqnarray}
&\displaystyle 
\frac{\partial F_{i}\left(
t,\mathbf{r}_{i},\mathbf{v}_{i}\right)} {\partial
t}+v_{ik}\frac{\partial F_{i}\left(
t,\mathbf{r}_{i},\mathbf{v}_{i}\right)} {\partial
r_{ik}}-\left(v_{il}+ V_{il}\left(
\mathbf{r}_{i}, t\right)\right)
\nonumber 
\\
 &\displaystyle
\times\int\limits^{t}_{t_{(0)}}\frac{\partial \hat{V}_{ik}\left(
\hat{\mathbf{r}}, \hat{t}_{1}\right)} {\partial \hat{r}_{il}}d\hat{t}_{1}\frac{\partial
F_{i}\left( t,\mathbf{r}_{i},\mathbf{v}_{i}\right)}{\partial r_{ik}}
\nonumber 
\\
 &\displaystyle
-v_{il}\frac{\partial \hat{V}_{ik} \left(\hat{\mathbf{r}},
\hat{t}_{1}\right)}{\partial \hat{r}_{il}}\frac{\partial F\left(
t,\mathbf{r}_{i}, \mathbf{v}_{i}\right)}{\partial
v_{ik}}+\frac{e_{i}}{m_{i}c}\left[\mathbf{v}_{i}
\times\mathbf{B}\right]\frac{\partial
F_{i}}{\partial\mathbf{v}_{i}}
\nonumber 
\\
 &\displaystyle
-\left\{ \left[\frac{\partial V_{ik}}{\partial\hat{t}}+V_{il}\frac{\partial V_{ik}}{\partial \hat{r}_{l}}
-\frac{e_{i}}{m_{i}}\left(\mathbf{E}_{0}\left(\mathbf{\hat{r}}, t\right)
+\frac{1}{c}\left[\hat{\mathbf{V}}_{i}\times\mathbf{B}\right]\right)_{k} \right]\right. 
\nonumber 
\\ 
&\displaystyle
\left.-\frac{e_{i}}{m_{i}}\frac{\partial\varphi\left(\hat{\mathbf{r}},\hat{t}\right)}{\partial
\hat{r}_{ik}}\right\}\frac{\partial F_{i}\left( t,
\mathbf{r}_{i},\mathbf{v}_{i}\right)}{\partial
v_{ik}} =0,\label{6}
\end{eqnarray}
We define the velocity $\mathbf{V}_{i}\left(\mathbf{r},t\right)$  as a function for which the square brackets in 
Eq. (\ref{6}) vanishes, 
\begin{eqnarray}
&\displaystyle 
\frac{\partial V_{ik}}{\partial\hat{t}}+V_{il}\frac{\partial V_{ik}}{\partial \hat{r}_{l}}
=\frac{e_{i}}{m_{i}}\left(\mathbf{E}_{0}\left(\mathbf{\hat{r}}, t\right)
+\frac{1}{c}\left[\hat{\mathbf{V}}_{i}\times\mathbf{B}\right]\right)_{k}.
\label{7}
\end{eqnarray}
The characteristic equations to first order partial differential equation (\ref{7}),
\begin{eqnarray}
&\displaystyle 
dt=\frac{d\hat{r}_{l}}{V_{il}}= \frac{d V_{ik}}{\frac{e_{i}}{m_{i}}\left(\mathbf{E}_{0}\left(\mathbf{\hat{r}}, t\right)
+\frac{1}{c}\left[\hat{\mathbf{V}}_{i}\times\mathbf{B}\right]\right)_{k}},\label{8}
\end{eqnarray}
display that the radius vector $\mathbf{r}_{i}$, determined by Eq. (\ref{4}) is the integral of Eq. (\ref{7}),
\begin{eqnarray}
&\displaystyle  \mathbf{\hat{r}}= \mathbf{r}_{i}+ \int\limits^{t}
_{t_{(0)}}\mathbf{V}_{i}\left(\mathbf{\hat{r}}, t_{1}\right)dt_{1}.
\label{9}
\end{eqnarray}
We obtain the solutions to nonlinear Eqs. (\ref{8}) for ion velocity components $V_{ix}\left(\mathbf{\hat{r}}, t\right)$ 
and $V_{iy}\left(\mathbf{\hat{r}}, t\right)$ in the limit $\mathbf{k}_{0}\mathbf{R}_{i}\left(\mathbf{\hat{r}}, t\right)= \mathbf{k}_{0}\int 
\mathbf{V}_{i}dt \sim \mathbf{k}_{0}\mathbf{V}_{i}/\omega_{0}\ll 1$, which corresponds to small displacement of 
ion in the fast wave compared with
fast wave wavelength $k^{-1}_{0}$. By integration by parts of Eq. (\ref{8}) for $V_{ik}$, a power series expansion 
in powers of $\left| \mathbf{k}_{0}\mathbf{V}
_{i}\left(\mathbf{r}_{i}, t\right)/\left(\omega_{0}\pm\omega_{ci}\right)\right| <1$ forms. We obtain on this way that
\begin{eqnarray}
&\displaystyle 
V_{ix}\left(\mathbf{\hat{r}}, t\right)=\frac{e}{2m_{i}}\frac{\left(E_{0x}+E_{0y}\right)}{\left(\omega_{0}+\omega_{ci}\right)}
\left(1-\frac{\mathbf{k}_{0}\mathbf{V}_{i}\left(\mathbf{r}_{i}, 
	t\right)}{\left(\omega_{0}+\omega_{ci}\right)}\right)
\nonumber 
\\
&\displaystyle
\times
\sin\left(\omega_{0}t - 
\mathbf{k}_{0}\left(\mathbf{r}_{i}+\mathbf{R}_{i}\left(\mathbf{r}_{i}, 
t\right)\right)\right)
\label{10}
\\
&\displaystyle
+\frac{e}{2m_{i}}\frac{\left(E_{0x}-E_{0y}\right)}{\left(\omega_{0}-\omega_{ci}\right)}
\left(1-\frac{\mathbf{k}_{0}\mathbf{V}_{i}\left(\mathbf{r}_{i}, 
	t\right)}{\left(\omega_{0}-\omega_{ci}\right)}\right)
\nonumber 
\\
&\displaystyle
\times
\sin\left(\omega_{0}t - \mathbf{k}_{0}\left(\mathbf{r}_{i}+
\mathbf{R}_{i}\left(\mathbf{r}_{i}, 
t\right)\right)\right)
+O\left[\left(\frac{\mathbf{k}_{0}\mathbf{V}_{i}\left(\mathbf{r}_{i}, t\right)}{\left(\omega_{0}\pm\omega_{ci}\right)}\right)^{2}\right],
\nonumber 
\end{eqnarray}
and
\begin{eqnarray}
&\displaystyle 
V_{iy}\left(\mathbf{\hat{r}}, t\right)=-\frac{e}{2m_{i}}\frac{\left(E_{0x}+E_{0y}\right)}{\left(\omega_{0}+\omega_{ci}\right)}
\left(1-\frac{\mathbf{k}_{0}\mathbf{V}_{i}\left(\mathbf{r}_{i}, 
	t\right)}{\left(\omega_{0}+\omega_{ci}\right)}\right)
\nonumber 
\\
&\displaystyle
\times
\cos\left(\omega_{0}t - 
\mathbf{k}_{0}\left(\mathbf{r}_{i}+\mathbf{R}_{i}\left(\mathbf{r}_{i}, 
t\right)\right)\right)
\label{11}
\\
&\displaystyle
+\frac{e}{2m_{i}}\frac{\left(E_{0x}-E_{0y}\right)}{\left(\omega_{0}-\omega_{ci}\right)}
\left(1-\frac{\mathbf{k}_{0}\mathbf{V}_{i}\left(\mathbf{r}_{i}, 
	t\right)}{\left(\omega_{0}-\omega_{ci}\right)}\right)
\nonumber 
\\
&\displaystyle
\times
\cos\left(\omega_{0}t - \mathbf{k}_{0}\left(\mathbf{r}_{i}+
\mathbf{R}_{i}\left(\mathbf{r}_{i}, 
t\right)\right)\right)
+O\left[\left(\frac{\mathbf{k}_{0}\mathbf{V}_{i}\left(\mathbf{r}_{i}, t\right)}{\left(\omega_{0}\pm\omega_{ci}\right)}\right)^{2}\right].
\nonumber 
\end{eqnarray}
The ion displacement $|\mathbf{R}_{i}\left(\mathbf{r}_{i}, t\right)|$ in the electric field of the fast wave is usually much less than 
the wavelength $k^{-1}_{0}$ of the fast wave, however it may be comparable with the wavelength of the unstable perturbations. 
Neglecting by the terms of the order of  $O\left(\mathbf{k}_{0}\mathbf{R}_{i}\left(\mathbf{r}_{i}, t\right)\right)\ll 1$ in Eqs. (\ref{9})-(\ref{11}), 
we obtain  that 
\begin{eqnarray}
&\displaystyle 
V_{ix}\left(\mathbf{\hat{r}}, t\right)=\frac{e}{2m_{i}}\frac{\left(E_{0x}+E_{0y}\right)}{\left(\omega_{0}+\omega_{ci}\right)}\sin\left(\omega_{0}t - 
\mathbf{k}_{0}\mathbf{r}_{i}\right)
\nonumber
\\
&\displaystyle
+\frac{e}{2m_{i}}\frac{\left(E_{0x}-E_{0y}\right)}{\left(\omega_{0}-\omega_{ci}\right)}\sin\left(\omega_{0}t - \mathbf{k}_{0}\mathbf{r}_{i}\right),
\label{12}
\\
&\displaystyle 
V_{iy}\left(\mathbf{\hat{r}}, t\right)=-\frac{e}{2m_{i}}\frac{\left(E_{0x}+E_{0y}\right)}{\left(\omega_{0}+\omega_{ci}\right)}\cos\left(\omega_{0}t - 
\mathbf{k}_{0}\mathbf{r}_{i}\right)
\nonumber 
\\
&\displaystyle
+\frac{e}{2m_{i}}\frac{\left(E_{0x}-E_{0y}\right)}{\left(\omega_{0}-\omega_{ci}\right)}\cos\left(\omega_{0}t - \mathbf{k}_{0}\mathbf{r}_{i}\right),
\label{13}
\\
&\displaystyle 
\hat{x}=x_{i}-\frac{e}{2m_{i}\omega_{0}}\frac{\left(E_{0x}+E_{0y}\right)}{\left(\omega_{0}+\omega_{ci}\right)}\cos\left(\omega_{0}t - 
\mathbf{k}_{0}\mathbf{r}_{i}\right)
\nonumber 
\\
&\displaystyle
-\frac{e}{2m_{i}\omega_{0}}\frac{\left(E_{0x}-E_{0y}\right)}{\left(\omega_{0}-\omega_{ci}\right)}\cos\left(\omega_{0}t - 
\mathbf{k}_{0}\mathbf{r}_{i}\right),
\label{14}
\end{eqnarray}
and
\begin{eqnarray}
&\displaystyle 
\hat{y}=y_{i}-\frac{e}{2m_{i}\omega_{0}}\frac{\left(E_{0x}+E_{0y}\right)}{\left(\omega_{0}+\omega_{ci}\right)}\sin\left(\omega_{0}t - 
\mathbf{k}_{0}\mathbf{r}_{i}\right)
\nonumber 
\\
&\displaystyle
+\frac{e}{2m_{i}\omega_{0}}\frac{\left(E_{0x}-E_{0y}\right)}{\left(\omega_{0}-\omega_{ci}\right)}\sin\left(\omega_{0}t - \mathbf{k}_{0}\mathbf{r}_{i}
\right).
\label{15}
\end{eqnarray}
Eqs. (\ref{12}) -- (\ref{15}) are the simplest results of the application of the local approximation, instead of the application of the approximation $\mathbf{k}_{0}=0$ of the spatially uniform FW, to the equations of the ion motion in the the spatially inhomogeneous electromagnetic wave.
In this approximation, where velocity $\mathbf{v}_{i}$ and coordinates $x_{i}, y_{i}$ are determined by Eqs. (\ref{4}) and (\ref{12}) -- (\ref{15}) in the 
ion  oscillating frame, the Vlasov equation (\ref{6}), in which the terms of the order of $O\left(|\mathbf{k}_{0}\mathbf{R}_{i}\left(\mathbf{r}_{i}, t
\right)|\right)\ll 1$ are neglected and the expression in the square brackets equal to zero, has a form 
\begin{eqnarray}
&\displaystyle 
\frac{\partial F_{i}\left(
t,\mathbf{r}_{i},\mathbf{v}_{i}\right)} {\partial
t}+\mathbf{v}_{i}\frac{\partial F_{i}\left(
t,\mathbf{r}_{i},\mathbf{v}_{i}\right)} {\partial
\mathbf{r}_{i}}
\nonumber 
\\
&\displaystyle
+\frac{e_{i}}{m_{i}}\left(-\nabla\varphi\left(\mathbf{r}_{i}, t\right)+ \frac{1}{c}\left[\mathbf{v}_{i}
\times\mathbf{B}\right]\right)\frac{\partial F_{i}}{\partial\mathbf{v}_{i}}=0.
\label{16}
\end{eqnarray}
Equation (\ref{16}) reveals that the  Vlasov equation for the ion distribution function in the variables 
$\mathbf{r}_{i}$,  $\mathbf{v}_{i}$
has a form as for uniform 
plasma without the external electromagnetic wave. The equilibrium ion distribution function 
$F_{i0}\left(t,\mathbf{r}_{i},
\mathbf{v}_{i}\right)$ in the variables $\mathbf{r}_{i}$,  $\mathbf{v}_{i}$ is determined by Eq. (\ref{16}) 
with $\varphi\left(\mathbf{r}_{i}, t\right)=0$ 
and may be assumed to be the Maxwellian distribution,
\begin{eqnarray}
&\displaystyle
F_{i0}\left(t,\mathbf{r}_{i},\mathbf{v}_{i}\right)=\frac{n_{0i}}{\left(2\pi v^{2}_{Ti}\right)^{3/2}}
\exp \left(-\frac{v^{2}_{i}}{2v^{2}_{Ti}}\right),
\label{17}
\end{eqnarray}
where $n_{0i}$ is the equilibrium ion density and $v^{2}_{Ti}=T_{i}/m_{i}$ is the ion thermal velocity.

All results presented above for the ion plasma component are completely applicable to the electron component. 
The Vlasov equation for the electron distribution function $F_{e}\left(t,\mathbf{r}_{e},\mathbf{v}_{e}\right)$
in the variables $\mathbf{r}_{e}$,  $\mathbf{v}_{e}$, determined 
by the relations (\ref{10}) -- (\ref{15}) with species index $i$ replaced by $e$, has a form of 
Eq. (\ref{16}) in which, however, 
the perturbed potential $\varphi$ is now the function of the coordinate $\mathbf{r}_{e}$ instead of $\mathbf{r}_{i}$.

\section{Electrostatic IC parametric instabilities driven by FW}\label{sec3}
In this section, we obtain the basic equations for the IC parametric instabilities for plasma
parameters consistent with cold SOL plasma near the FW antenna region in tokamak, 
where the oscillatory velocity of ions and electrons may be commensurate
with their thermal velocities.

\subsection{Ion and electron oscillating modes and their interactions}\label{sec3.1}
It follows from Eq. (\ref{16}) that with 
variables $\mathbf{r}_{i}$, $\mathbf{v}_{i}$ determined in the ion frame of references which 
oscillates with velocity $\mathbf{V}_{i}$ of ions in the crossed electric and magnetic fields of FW, the 
Vlasov equation for $f_{i}\left(t, \mathbf{r}_{i},\mathbf{v}_{i}\right)$, 
which contains the potential $\varphi\left(\mathbf{r}_{i}, t\right)$, 
has a form as for a plasma without FW fields. With leading center coordinates 
$X_{i}$, $Y_{i}$, $z_{i}$ determined in the ion oscillating frame by the relations 
\begin{eqnarray}
&\displaystyle 
x_{i}=X_{i}-\frac{v_{i\bot}}{\omega_{ci}}\sin \left(\phi_{1}-\omega_{ci}t\right), 
\nonumber 
\\
&\displaystyle
y_{i}=Y_{i}+\frac{v_{i\bot}}{\omega_{ci}}
\cos \left(\phi_{1}-
\omega_{ci}t\right),
\label{18}
\end{eqnarray}
where \quad
\begin{eqnarray}
&\displaystyle 
\phi=\phi_{1}-\omega_{ci}t, \quad v_{ix}=v_{i\bot}\cos \phi, \nonumber 
\\
&\displaystyle
v_{iy}=v_{i\bot}\sin \phi ,
\label{19}
\end{eqnarray}
the Vlasov equation for $f^{(0)}_{i}$ has a form
\begin{eqnarray}
&\displaystyle 
\frac{\partial f_{i}}{\partial
t}+\frac{e_{i}}{m_{i}\omega_{ci}}\left(\frac{\partial\varphi}{\partial X_{i}} \frac{\partial
f_{i}} {\partial Y_{i}}-\frac{\partial\varphi}{\partial
Y_{i}} \frac{\partial f_{i}} {\partial X_{i}}\right)
\nonumber 
\\
&\displaystyle
+\frac{e_{i}}{m_{i}}\frac{\omega_{ci}}{v_{i\bot}}
\left(\frac{\partial\varphi}{\partial \phi_{1}} \frac{\partial
f_{i}} {\partial v_{i\bot}}-\frac{\partial\varphi}{\partial
v_{i\bot}}\frac{\partial
f_{i}} {\partial \phi_{1}}\right)-\frac{e_{i}}{m_{i}}\frac{\partial\varphi}{\partial
z_{i}} \frac{\partial f_{i}}{\partial v_{iz}}
\nonumber 
\\
&\displaystyle
=- \frac{e_{i}}{m_{i}}\frac{\omega_{ci}}{v_{i\bot}}\frac{\partial\varphi}{\partial\phi_{1}}
\frac{\partial F_{i0}}{\partial
v_{i\bot}}+\frac{e_{i}}{m_{i}}\frac{\partial\varphi}{\partial z_{i}}
\frac{\partial F_{i0}}{\partial v_{iz}}. 
\label{20}
\end{eqnarray}
Instead of the linearisation of Eq. (\ref{20}) as a routine initial step to the solution of 
this nonlinear equation, we solve Eq. (\ref{20}) employing the 
procedure of the "renormalized linearisation" \cite{Mikhailenko3} which provides the 
inclusion to the  
derived linear solution for $f_{i}$ the averaged nonlinear effect of the scattering of ions 
by the ensemble of the IC waves\cite{Dum,Benford}. This effect, which is interpreted as the 
"effective 
collision" of ions with IC turbulence or as the effect of the broadening of the IC 
resonances, is the dominant one in the saturation of the IC kinetic parametric instabilities 
and anomalous heating of the non resonant ions by the IC turbulence\cite{Mikhailenko3}. 

The procedure of the "renormalized linearisation" 
consists in the transformation of the leading center 
coordinates $X_{i}$, $Y_{i}$ 
and velocity  coordinates $v_{i\bot}$, $\phi_{1}$, $v_{iz}$ in Eq. (\ref{20}) to new 
coordinates, which account for in the explicit form the distortion of 
the ion orbits by the electrostatic turbulence. 
Assuming  that the ion orbit disturbances $\delta X_{i}$, $\delta Y_{i}$,
 $\delta v_{i\bot}$, $\delta\phi$, $\delta v_{iz}$ are sufficiently small, these coordinates are\cite{Mikhailenko3}
\begin{eqnarray}
&\displaystyle 
X_{i}=\bar{X}_{i}+\delta X_{i}  \quad \delta
X_{i}=-\frac{e}{m_{i}\omega_{ci}}\int\limits^{t}\frac{\partial\varphi}
{\partial \bar{Y}_{i}}dt_{1},
\label{21}
\\
&\displaystyle 
Y_{i}=\bar{Y}_{i}+\delta Y_{i},  \quad \delta
Y_{i}=\frac{e}{m_{i}\omega_{ci}}\int\limits^{t}\frac{\partial\varphi}
{\partial \bar{X}_{i}}dt_{1},
\label{22}
\\
&\displaystyle 
v_{i\bot}=\bar{v}_{i\bot}+\delta v_{i\bot},  \quad 
\delta v_{i\bot}=
\frac{e}{m_{i}}\frac{\omega_{ci}}{v_{i\bot}}
\int\limits^{t}\frac{\partial\varphi}{\partial
\bar{\phi}}dt_{1},
\label{23}
\\
&\displaystyle 
\phi=\bar{\phi}+\delta\phi,    \quad
\delta\phi=-\frac{e}{m_{i}}\frac{\omega_{ci}}{v_{i\bot}}
\int\limits^{t}\frac{\partial\varphi}{\partial
\bar{v}_{i\bot}}dt_{1},
\label{24}
\\
&\displaystyle 
v_{iz}=\bar{v}_{iz}+\delta v_{iz}, \quad \delta v_{iz}
=-\frac{e}{m}_{i}\int\limits^{t}_{0}\frac{\partial\varphi}{\partial
\bar{z}_{i}}dt_{1}.
\label{25}
\end{eqnarray}
in which $\bar{X}_{i}$, $\bar{Y}_{i}$, $\bar{v}_{i\bot}$, $\bar{\phi}$, $\bar{v}_{iz}$ 
are the integrals of the system of equations for the characteristics to Eq. (\ref{20}). 
The perturbed electrostatic potential $\varphi$ is presented
in variables $\bar{X}_{i}$, $\bar{Y}_{i}$, $\bar{v}_{i\bot}$, $\bar{\phi}$, $\bar{v}_{iz}$ and $\delta X_{i}$,
$\delta Y_{i}$, $\delta v_{i\bot}$, $\delta\phi$, $\delta v_{iz}$ in a form
\begin{eqnarray}
& \displaystyle
 \varphi\left(\mathbf{r}_{i},t \right)=\int
d\mathbf{k}d\omega
\varphi\left(\mathbf{k}_{i},\omega \right)
\nonumber
\\
&\displaystyle 
\times\exp\Big[-i\omega t+ik_{ix}x_{i} +ik_{iy}y_{i}+ik_{iz}z_{i}\Big] 
\nonumber
\\
&\displaystyle 
=\sum_{n=-\infty}^{\infty}\int
d\mathbf{k}d\omega
\varphi\left(\mathbf{k}_{i},\omega \right)J_{n}\left(
\frac{k_{i\bot}\bar{v}_{i\bot}}{\omega_{ci}}\right)
\nonumber\\
&\displaystyle 
\times\exp\left(i\mathbf{k}_{i}\delta\mathbf{r}_{i}\left(t\right)\right)
\exp\Big[-i\omega t+ik_{ix}\bar{X}_{i} 
\nonumber\\ &
\displaystyle+ik_{iy}\bar{Y}_{i}+ik_{iz}\left(\bar{z}_{i} +\bar{v}_{iz}t\right)-in\left(\bar{\phi}-\omega_{ci}t-\theta \right)\Big],
\label{26}
\end{eqnarray}
where $J_{n}$ is the Bessel function of the order $n$. The subscript $i$ in $\varphi_{i}$ and of $\mathbf{k}_{i}$ 
indicates that $\varphi_{i}=
\varphi\left(\mathbf{r}_{i}, t\right)$ and that $\mathbf{k}_{i}$ is the wave vector conjugate 
to $\mathbf{r}_{i}$. The nonlinear phase shift $\mathbf{k}_{i}\delta\mathbf{r}_{i}\left(t\right)$, 
resulted from the perturbations of the ions orbits, 
\begin{eqnarray}
& \displaystyle \mathbf{k}\delta\mathbf{r}_{i}\left(t\right)=k_{x}\delta X_{i}+k_{y}\delta Y_{i}
\nonumber\\
&\displaystyle
+k_{z}\int\limits^{t}\delta v_{iz}\left(\tau \right)
d\tau 
\nonumber\\ & \displaystyle
-\frac{k_{\bot}\delta v_{i\bot}}{\omega_{ci}}\sin
\left(\phi-\theta\right)
-\frac{k_{\bot}\bar{v}_{i\bot}}{\omega_{ci}}\cos
\left(\phi-\theta\right)\delta\phi, \label{27}
\end{eqnarray}
is included. In Eq.(\ref{27}) the terms of the second order over
$\delta X_{i}$, $\delta Y_{i}$, $\delta v_{i\perp}$, $\delta v_{iz}$ and
$\delta\phi$ are omitted. It was obtained in Ref. \cite{Mikhailenko3} that with new variables 
$t$, $\bar{X}$, $\bar{Y}$, $\bar{z}_{1}$, $\bar{v}_{\bot}$, $\bar{\phi}_{1}$, $\bar{v}_{z}$, determined above by the  
relations (\ref{21}) -- (\ref{25}), the nonlinearities of the second order in the equation for
$f_{i}\left(t_{1}, \bar{X}_{i}, \bar{Y}_{i}, \bar{z}_{1}, \bar{v}_{i\bot}, \bar{\phi}_{1}, \bar{v}_{iz}\right)$ 
convert into the nonlinearities of the third order 
with respect to the potential $\varphi$ in the left-hand-side of Eq. (\ref{20}).
Neglecting by these nonliearities, we obtain the linear equation for $f_{i}$ with known $F_{i0}$
\begin{eqnarray}
& \displaystyle 
\frac{\partial f_{i}}{\partial t} =
\frac{e}{m}_{i}\left[-\frac{\omega_{ci}}{\bar{v}_{i\bot}}
\frac{\partial\varphi}{\partial \bar{\phi}_{1}} \frac{\partial
F_{i0}}{\partial \bar{v}_{i\bot}} +\frac{\partial\varphi}{\partial z_{1}}
\frac{\partial F_{i0}}{\partial \bar{v}_{iz}}\right]  
\label{28}
\end{eqnarray}
with the solution 
\begin{eqnarray}
& \displaystyle f_{i}\left(t, \bar{X}, \bar{Y}, \bar{z}_{1}, \bar{v}_{\bot}, \bar{\phi}_{1}, \bar{v}_{z}\right) 
\nonumber
\\
&\displaystyle 
=\frac{e}{m}_{i}\int\limits^{t}\left[-\frac{\omega_{ci}}{\bar{v}_{\bot}}
\frac{\partial\varphi}{\partial \bar{\phi}_{1}} \frac{\partial
F_{i0}}{\partial \bar{v}_{\bot}} +\frac{\partial\varphi}{\partial z_{1}}
\frac{\partial F_{i0}}{\partial \bar{v}_{z}} \right] dt'. \label{29}
\end{eqnarray}
As in the conventional renormalized theory\cite{Dum}, we account for in Eq. (\ref{29}) the average effect of the
perturbations of ions orbits. For the accounting for the averaging effect of the ion scattering by the electrostatic 
turbulent field we use the simplified approximation that the  scattering of particles by plasma turbulence is a 
Gaussian process, for which the relation \cite{Dum},
\begin{eqnarray}
& \displaystyle
\left\langle e^{-i\mathbf{k}_{\bot}\left(\delta\mathbf{r}\left(t\right)-\delta\mathbf{r}\left(t_{1}\right)\right)}\right\rangle
\nonumber
\\
&\displaystyle 
\simeq e^{-\frac{1}{2}\left\langle\left(\mathbf{k}_{\bot}\delta\mathbf{r}
\left(t-t_{1}\right)\right)^{2}\right\rangle}=e^{-C_{i}\left(t-t_{1}\right)},\label{30}
\end{eqnarray}
holds. The coefficient $C_{i}$ in Eq. (\ref{30}) is \cite{Mikhailenko3} 
\begin{eqnarray}
& \displaystyle C_{i}=\frac{e^{2}}{2m^{2}_{i}}Re \sum\limits_{n=-\infty}^{\infty}\int d\mathbf{k}_{1}|
\varphi\left(\mathbf{k}_{1\bot}\right)|^{2}
\mathcal{F}_{i}\left( k_{\bot},k_{1\bot}\right)e^{-k_{\bot}^{2}\rho^{2}_{i}}
\nonumber
\\
&\displaystyle 
\times
\sqrt{\frac{\pi}{2}}\frac{1}{k_{z}v_{Ti}}
W\left(\frac{\omega-n_{1}\omega_{ci}}{\sqrt{2}k_{z}v_{Ti}}\right),
\label{31}
\end{eqnarray}
and
\begin{eqnarray}
& \displaystyle \mathcal{F}_{i}\left( k_{\bot},k_{1\bot}\right)=\frac{2}{\omega^{2}_{ci}}\left(k_{x}k_{1y}-k_{y}k_{1x}\right)^{2}
I_{n}\left(k^{2}_{\bot}\rho^{2}_{i}\right)
\nonumber\\ 
& \displaystyle
+\frac{1}{2}\frac{k^{2}_{\bot}k^{2}_{1\bot}}{\omega^{2}_{ci}}\Big(I_{n+1}\left(k^{2}_{\bot}\rho^{2}_{i}\right)
+I_{n-1}\left(k^{2}_{\bot}\rho^{2}_{i}\right)\Big).
\label{32}
\end{eqnarray}
The solution (\ref{29}) for  $f_{i}$, Fourier transformed over $\mathbf{r}_{i}$, 
which accounts for the average effect of the scattering of ions by the electrostatic turbulence, is calculated easily and is equal to
\begin{eqnarray}
&\displaystyle 
f_{i}\left(t,\mathbf{k}_{i}, v_{i\bot},\phi,v_{iz}\right)=\frac{ie}{m_{i}}\sum\limits_{n=-\infty}^{\infty}
\sum\limits_{n_{1}=-\infty}^{\infty} \int\limits_{t_{0}}^{t}dt_{1}\varphi\left(\mathbf{k}_{i}, t_{1}\right)
\nonumber  
\\
&\displaystyle
\times
\exp\Big[-ik_{iz}v_{iz}\left(t-t_{1}\right)-C_{i}\left(t-t_{1}\right)
\nonumber
\\ 
&\displaystyle
+in\left(
\phi_{1}-\omega_{ci}t-\theta\right)-in_{1}\left(
\phi_{1}-\omega_{ci}t_{1}-\theta\right) \Big]
\nonumber
\\ 
&\displaystyle
\times
J_{n}\left(\frac{k_{i\bot}v_{i\bot}}{\omega_{ci}}\right) J_{n_{1}}
\left(\frac{k_{i\bot}v_{i\bot}}{\omega_{ci}}\right)
\nonumber
\\ 
&\displaystyle
\times
\left[\frac{\omega_{ci}n_{1}}{v_{i\bot}}
\frac{\partial F_{i0}}{\partial v_{i\bot}}+ k_{iz}\frac{\partial
F_{i0}}{\partial v_{iz}}\right], 
\label{33}
\end{eqnarray}
where $t_{0}\geq 0$ is the initial time. For the Maxwellian distribution (\ref{17}) 
for $F_{i0}\left(t,\mathbf{r}_{i},\mathbf{v}_{i}\right)$, 
the perturbation of the ion density 
Fourier transformed over $\mathbf{r}_{i}$, $\delta n_{i}\left(\mathbf{k}_{i}, t\right)
=\int d\mathbf{v}_{i}f_{i}\left(\mathbf{v}_{i}, \mathbf{k}_{i}, t\right)$, is equal to
\begin{eqnarray}
& \displaystyle 
\delta n_{i}\left(\mathbf{k}_{i}, t\right)=i\frac{2\pi e_{i}}{m_{i}}
\sum^{\infty}_{n=-\infty}I_{n}\left(k^{2}_{\bot}\rho^{2}_{i}\right)e^{-k^{2}_{\bot}\rho^{2}_{i}}
\nonumber
\\
&\displaystyle 
\times
\int\limits^{t}_{t_{0}}dt_{1}
\varphi_{i}\left(\mathbf{k}
_{i}, t_{1}\right)
\left(-n\omega_{ci}+ik^{2}_{iz}v^{2}_{Ti}\left(t-t_{1}\right)\right)
\nonumber
\\ 
&\displaystyle
\times
e^{-in\omega_{ci}\left(t-t_{1}\right)-C_{i}\left(t-t_{1}\right)-
	\frac{1}{2}k^{2}_{iz}v^{2}_{Ti}\left(t-t_{1}\right)^{2}} 
\label{34}
\end{eqnarray}

The solution to the linearised Vlasov equation for the separate spatial Fourier harmonic of the perturbation $f_{e}$ 
of the electron distribution function 
$F_{e0}$ and the solution for the perturbed electron density $\delta n_{e}$, 
\begin{eqnarray}
& \displaystyle \delta n_{e}\left(\mathbf{k}_{e}, t\right)=-\frac{2\pi e}{m_{e}}
\int\limits^{t}_{t_{0}}dt_{1}\varphi_{e}\left(\mathbf{k}_{e}, t_{1}\right)
\nonumber
\\ 
&\displaystyle
\times
k^{2}_{ez}v^{2}_{Te}\left(t-t_{1}\right)
e^{-\frac{1}{2}k^{2}_{ez}v^{2}_{Te}\left(t-t_{1}\right)^{2}}, 
\label{35}
\end{eqnarray}
are derived in the electron oscillating frame as a functions of the Fourier transformed perturbed potential 
$\varphi_{e}\left(\mathbf{k}_{e}, 
t\right)=\int d\mathbf{r}_{e}\varphi_{e}\left(\mathbf{r}_{e}, t\right)e^{-i\mathbf{k}_{e}
\mathbf{r}_{e}}$. It is obvious that $\varphi_{e}\left(\mathbf{k}_{e}, t\right)$ and $
\varphi_{i}\left(\mathbf{k}_{i}, t\right)$ are different functions. The temporal evolution of the separate spatial harmonic 
of the potential 
$\varphi$  with Poisson equation (\ref{3}) may be investigated  in the ion frame as the 
equation for $\varphi_{i}\left(\mathbf{k}_{i},t\right)$ by the Fourier transform of Eq. 
(\ref{3}) over $\mathbf{r}_{i}$, 
\begin{eqnarray}
& \displaystyle 
k^{2}_{i}\varphi_{i}\left(\mathbf{k}_{i},t\right)
=4\pi e
\nonumber
\\
&\displaystyle 
\times
\left(\delta n_{i}\left(\mathbf{k}_{i}, t\right)
-\int d\mathbf{r}_{i}\delta n_{e}\left(\mathbf{r}_{e}, t\right)e^{-i\mathbf{k}_{i}\mathbf{r}_{i}}\right).
\label{36}
\end{eqnarray}
or as the equation for  $\varphi_{e}\left(\mathbf{k}_{e},t\right)$ by the Fourier transform of Eq. (\ref{3}) over $\mathbf{r}_{e}$. 
For deriving 
the Fourier transformed Poisson equation (\ref{3}) for $\varphi_{i}\left(\mathbf{k}_{i},t\right)$, the Fourier transform over 
$\mathbf{r}_{i}$ should be determined for $n_{e}
\left(\mathbf{r}_{e},t\right)$ as well as for potential $\varphi_{e}\left(\mathbf{r}_{e},t_{1} 
\right)$, which is included in the expression for $n_{e}
\left(\mathbf{r}_{e},t\right)$. With coordinates transform (\ref{4}) for the ions and with similar transforms for the electrons 
we obtain, that
\begin{eqnarray}
& \displaystyle 
\int d\mathbf{r}_{i}\delta n_{e}\left(\mathbf{r}_{e}, t\right)e^{-i\mathbf{k}_{i}\mathbf{r}_{i}}=\sum\limits_{m=-\infty}^{\infty}
J_{m}\left(a_{ei}\right)e^{im\left(\omega_{0}t+\delta\right)}
\nonumber\\
&\displaystyle 
\times
\delta n_{e}^{(e)}\left(\mathbf{k}_{i}-m\mathbf{k}_{0}, t\right),
\label{37}
\end{eqnarray}
where
\begin{widetext}
\begin{eqnarray}
& \displaystyle 
a_{ei}=\left\lbrace\left[\sum\limits_{\alpha=i, e}\frac{e_{\alpha}k_{iy}}{2m_{\alpha}\omega_{0}}\left(\frac{E_{0x}-E_{0y}}
{\omega_{0}-\omega_{c\alpha}} -\frac{E_{0x}+E_{0y}}{\omega_{0}+\omega_{c\alpha}}\right)  \right]^{2}
+\left[\sum\limits_{\alpha=i, e}\frac{e_{\alpha}k_{ix}}{2m_{\alpha}\omega_{0}}\left(\frac{E_{0x}+E_{0y}}
{\omega_{0}+\omega_{c\alpha}} +\frac{E_{0x}-E_{0y}}{\omega_{0}-\omega_{c\alpha}}\right) \right]^{2}  \right\rbrace ^{1/2},
\label{38}
\end{eqnarray}
\end{widetext}
and 
\begin{eqnarray}
& \displaystyle 
\tan \delta=\frac{\sum\limits_{\alpha=i, e}\frac{e_{\alpha}k_{iy}}{m_{\alpha}}\left(\frac{E_{0x}+E_{0y}}{\omega_{0}+
\omega_{c\alpha}}+\frac{E_{0x}-E_{0y}}
{\omega_{0}-\omega_{c\alpha}}\right)  
}{\sum\limits_{\alpha=i, e}\frac{e_{\alpha}k_{ix}}{m_{\alpha}}\left(\frac{E_{0x}+E_{0y}}
{\omega_{0}+\omega_{c\alpha}} -\frac{E_{0x}-E_{0y}}{\omega_{0}-\omega_{c\alpha}}\right)}.
\label{39}
\end{eqnarray}
It follows from Eqs. (\ref{14}) and (\ref{15}), that $a_{ei}\sim |\mathbf{k}_{i}\xi_{ie}| $, where $\xi_{ie}$ is the amplitude 
of the displacements of electrons relative ions in FW. 

The relation between the Fourier transform $\varphi_{e}\left( \mathbf{k}_{e}, t\right)$ of the potential $\varphi_{e}
\left( \mathbf{r}_{e}, t\right)$ over 
$\mathbf{r}_{e}$, involved in the expression 
for $\delta n_{e}\left(\mathbf{r}_{e}, t\right)$, and the Fourier transform $\varphi_{i}\left( \mathbf{k}_{i}, t\right)$ 
of the potential 
$\varphi_{i}\left( \mathbf{r}_{i}, t\right)$ over $\mathbf{r}_{i}$ when it is 
used in $\delta n_{e}$ instead of $\varphi_{e}\left( \mathbf{k}_{e}, t\right)$, is derived similar and is determined by 
the relation
\begin{eqnarray}
&\displaystyle \varphi_{e}\left(\mathbf{k}_{e},t_{1}\right)=\int d\mathbf{r}_{e}\varphi_{e}
\left(\mathbf{r}_{e},t_{1}\right)e^{-i\mathbf{k}_{e}\mathbf{r}_{e}}
\nonumber
\\ 
&\displaystyle
=\sum\limits_{p=-\infty}^{\infty}J_{p}\left(a_{ei}\right)e^{ip\left(\omega_{0}t_{1}+\delta\right)}
\nonumber 
\\
&\displaystyle
\times
\varphi_{i}\left(\mathbf{k}_{i}-\left(m-p\right)\mathbf{k}_{0}, t_{1}\right).
\label{40}
\end{eqnarray}
This relation follows from the identity $\varphi_{e}\left(\mathbf{r}_{e},t_{1}\right)=\varphi_{i}\left(\mathbf{r}_{i},t_{1}\right)$, 
and relation (\ref{37}). The Fourier transform over the time of the Poisson equation for the potential 
$\varphi_{i}\left(\mathbf{k}_{i}, t\right)$
gives the basic equation,
\begin{eqnarray}
&\displaystyle \varepsilon\left(\mathbf{k}_{i}, \omega\right)\varphi_{i}\left(\mathbf{k}_{i}, \omega\right)
+\sum\limits_{q\neq 0}\sum\limits_{m=-\infty}^{\infty}J_{m}\left(a_{ei}\right)
J_{m+q}\left(a_{ei}\right)e^{iq\delta}
\nonumber
\\ 
&\displaystyle
\times\varepsilon_{e}\left(\mathbf{k}_{i}-m\mathbf{k}_{0}, \omega-m\omega_{0}\right)
\nonumber
\\ 
&\displaystyle
\times
\varphi_{i}\left(\mathbf{k}_{i}+q\mathbf{k}_{0}, \omega+q\omega_{0}\right)=0,
\label{41}
\end{eqnarray}
where
\begin{eqnarray}
&\displaystyle \varepsilon\left(\mathbf{k}_{i}, \omega\right)=1+\varepsilon_{i}\left(\mathbf{k}_{i}, \hat{\omega}\right)
\nonumber
\\ 
&\displaystyle
+ \sum\limits_{m=-\infty}^{\infty}
J^{2}_{m}\left(a_{ei}\right)\varepsilon_{e}\left(\mathbf{k}_{i}-m\mathbf{k}_{0}, \omega-m\omega_{0}\right).
\label{42}
\end{eqnarray}
The functions $\varepsilon_{i}\left(\mathbf{k}_{i}, \hat{\omega}\right)$ and $\varepsilon_{e}\left(\mathbf{k}_{i}, \omega\right)$ in 
Eq. (\ref{42}) are the renormalized nonlinear dielectric permittivity of ions and the linear permittivity of electrons 
respectively which for 
the  Maxwellian distribution (\ref{17}) for ions and electrons are
\begin{eqnarray}
&\displaystyle \varepsilon_{i}\left(\mathbf{k}_{i}, \hat{\omega}\right)=\frac{1}{k^2\lambda _{Di }^2}\left[1 + i\sqrt {\frac{\pi }{2}}
\frac{\hat{\omega}}{k_{z} v_{Ti}}\sum\limits_{n = - \infty }^{\infty}
W\left(z_{in}\right)\right.
\nonumber
\\ 
&\displaystyle
\times 
I_{n}\left(k^{2}_{\bot}\rho^{2}_{i}\right)\exp\left(-k^{2}_{\bot}\rho^{2}_{i}\right)\Big], 
\label{43}
\\
&\displaystyle 
\varepsilon_{e}\left(\mathbf{k}_{i}, \omega-m\omega_{0}\right)
=\frac{1}{k^2\lambda _{De}^2}
\nonumber
\\ 
&\displaystyle
\times
\Big(1+i\sqrt{\pi}\left(z_{e}-m\zeta_{e}\right)W\left(z_{e}-m\zeta_{e} \right)\Big).
\label{44}
\end{eqnarray}
In Eqs. (\ref{43}), (\ref{44}), $\lambda_{Di(e)}$ is the ion (electron) Debye length, $\rho_{i}= v_{Ti}/\omega_{ci}$  
is the ion thermal Larmor radius, 
$I_{n}$ is the modified Bessel function of order $n$, \\ $W\left(z\right)=e^{-z^{2}}\left(1 +\left(2i/ \sqrt {\pi } \right)
\int\limits_{0}^{z} e^{t^{2}}dt 
\right)$ is the complex error function,  $z_{in} =\left(\hat{\omega}-n\omega_{ci}\right)/\sqrt{2}k_{z}v_{Ti}$, 
$z_{e} = \omega/\sqrt{2}k_{z}v_{Te}$, and $\zeta_{e} = \omega_{0}/\sqrt{2}k_{z}v_{Te}$. The renormalized frequency 
$\hat{\omega}=\omega+iC_{i}$, where 
$C_{i}$ is determined by Eq. (\ref{31}), accounts for the scattering of ions by the ion cyclotron turbulence.   

Equation (\ref{41}) describes the coupling of the fundamental mode $\varphi_{i}\left(\mathbf{k}_{i}, \omega\right)$ 
with their harmonics $\varphi_{i}
\left(\mathbf{k}_{i}+q\mathbf{k}_{0}, \omega+q\omega_{0}\right)$. The equation for the harmonic $\varphi_{i}\left(\mathbf{k}_{i
\bot}+t\mathbf{k}_{0\bot}, k_{iz}, \omega+t\omega_{0}\right)$ is derived from Eq. (\ref{41}) by changing $\omega$ on $\omega+t\omega_{0}$, 
and $\mathbf{k}
_{i}$ on $\mathbf{k}_{i}+t\mathbf{k}_{0}$,
\begin{eqnarray}
&\displaystyle 
\varphi_{i}\left(\mathbf{k}_{i}+t\mathbf{k}_{0}, \omega+t\omega_{0}\right)
=-\frac{1}{\varepsilon\left(\mathbf{k}_{i}+t\mathbf{k}_{0}, \omega+t\omega_{0}\right)}
\nonumber
\\ 
&\displaystyle
\times
\sum\limits_{q_{1}\neq 0}\sum\limits_{m_{1}
=-\infty}^{\infty}J_{m_{1}}\left(a_{ei}\right)J_{m_{1}+q_{1}}\left(a_{ei}\right)e^{iq_{1}\delta}
\nonumber
\\ 
&\displaystyle
\times\varepsilon_{e}\left(\mathbf{k}_{i}-\left(m_{1}-t\right)\mathbf{k}_{0}, \omega-\left(m_{1}-t\right)
\omega_{0}\right)
\nonumber
\\ 
&\displaystyle
\times
\varphi_{i}\left(\mathbf{k}_{i\bot}+ \left(q_{1}+t\right)\mathbf{k}_{0}, \omega+\left(q_{1}+t\right)\omega_{0}\right),
\label{45}
\end{eqnarray}
Employing Eq. (\ref{45}) with $t=q$ in Eq. (\ref{41}), we obtain the equation
\begin{eqnarray}
&\displaystyle
 \hat{\varepsilon}\left(\mathbf{k}_{i}, \omega\right)\varphi_{i}\left(\mathbf{k}, \omega\right)
-\sum\limits_{q\neq 0}\sum\limits_{q_{1}\neq 0, (q_{1}\neq q)}\sum\limits_{m=-\infty}^{\infty}
\sum\limits_{m_{1}=-\infty}^{\infty}
\nonumber 
\\
&\displaystyle
\times
J_{m}\left(a_{ei}
\right)J_{m_{1}}\left(a_{ei}\right)
\nonumber
\\ 
&\displaystyle
\times
J_{m+q}\left(a_{ei}\right)J_{m_{1}+q_{1}}\left(a_{ei}\right)e^{i\left(q+q_{1}\right)\delta}
\nonumber
\\ 
&\displaystyle
\times\frac{\varepsilon_{e}\left(\mathbf{k}_{i}-m\mathbf{k}_{0}, \omega-m\omega_{0}\right)}
{\varepsilon\left(\mathbf{k}_{i}+q\mathbf{k}_{0}, \omega+q\omega_{0}\right)}
\nonumber 
\\
&\displaystyle
\times
\varepsilon_{e}\left(\mathbf{k}_{i}-\left(m_{1}-q\right)
\mathbf{k}_{0}, \omega-\left(m_{1}-q\right)\omega_{0}\right)
\nonumber
\\ 
&\displaystyle
\times
\varphi_{i}\left(\mathbf{k}_{i}+\left(q+q_{1}\right)\mathbf{k}_{0}, \omega+\left(q+q_{1}\right)\omega_{0}\right)=0,
\label{46}
\end{eqnarray}
where
\begin{eqnarray}
&\displaystyle \hat{\varepsilon}\left(\mathbf{k}_{i}, \omega\right)= \varepsilon\left(\mathbf{k}_{i}, \omega\right)
-\sum\limits_{q\neq 0}\sum\limits_{m=-\infty}^{\infty}\sum\limits_{m_{1}=-\infty}^{\infty}
\nonumber 
\\
&\displaystyle
\times
J_{m}\left(a_{ei}\right)J_{m_{1}}\left(a_{ei}\right)J_{m+q}
\left(a_{ei}\right)J_{m_{1}-q}\left(a_{ei}\right)
\nonumber
\\ 
&\displaystyle
\times\frac{\varepsilon_{e}\left(\mathbf{k}_{i}-m\mathbf{k}_{0}, \omega-m\omega_{0}\right)
}{\varepsilon\left(\mathbf{k}_{i}+q\mathbf{k}_{0}, \omega+q\omega_{0}\right)}
\nonumber 
\\
&\displaystyle
\times
\varepsilon_{e}\left(\mathbf{k}_{i}-\left(m_{1}-q\right)
\mathbf{k}_{0}, \omega-\left(m_{1}-q\right)\omega_{0}
\right).
\label{47}
\end{eqnarray}
This process of the separation of the term with potential $\varphi_{i}\left(\mathbf{k}, 
\omega\right)$ of the fundamental mode in 
Eq. (\ref{46}) may be repeated ad infinitum \cite{Mikhailenko4}. At the n-th step of the 
iteration, the terms of the order of $O
\left(\left(\varepsilon_{e}/\varepsilon\right)^{n}\right)$ are added to $\hat{\varepsilon}
\left(\mathbf{k}_{i}, \omega\right)$. 
This process converges when $\left| \varepsilon_{e}/\varepsilon\right| < 1$ and, it is 
important to note, the product of the growing number of 
Bessel functions is much 
less than unity and supplies the convergence of the procedure even when $\left| 
\varepsilon_{e}/\varepsilon\right| \sim 1$. If we neglect by the next 
iteration step, which adds to $\hat{\varepsilon}$, determined by Eq. (\ref{47}), 
the terms of the order of $\left| \varepsilon_{e}/\varepsilon\right| ^{3}$ multiplied 
on the product of eight Bessel functions, we obtain the equation 
\begin{eqnarray}
&\displaystyle \hat{\varepsilon}\left(\mathbf{k}_{i}, \omega\right)=0 
\label{48}
\end{eqnarray}
which is the general form of the dispersion equation for the plasma in FW, the particular 
cases of which compose the contemporary theory of parametric instabilities of plasmas in FW.

\subsection{IC kinetic parametric instability.}\label{sec3.2}
The investigation of the dispersion equation (\ref{48}) we begin with considering the simplest case which corresponds to the IC kinetic 
parametric instability\cite{Mikhailenko1, Mikhailenko4}.  
The dispersion properties of this instability are determined by the solution of the equation 
\begin{eqnarray}
&\displaystyle \varepsilon\left(\mathbf{k}_{i}, \omega\right)=0.
\label{49}
\end{eqnarray}
It determines the stability properties of the fundamental mode $\varphi_{i}\left(\mathbf{k}
_{i}, \omega\right)$ without accounting for the effects of the coupling  this mode with 
harmonics $\varphi_{i}\left(\mathbf{k}_{i}+q\mathbf{k}_{0}, \omega+q\omega_{0}\right)$ (see 
Eq. (\ref{41})). In Eq. (\ref{49}),  $\varepsilon\left(\mathbf{k}_{i}, \omega\right)$ is 
given by Eq. (\ref{42}). This instability is driven by the oscillation current formed 
by the relative oscillatory motion of ions and electrons. The linear and the nonlinear theory 
of this instability was investigated in detail in Ref. \cite{Mikhailenko4}. Here we present 
only a short summary of the 
derived results for the comparing them with the corresponding results for the IC quasimode 
decay instability considered below in Subsec. \ref{sec3.2} and Sec. \ref{sec4}.

It follows from Eq. (\ref{40}) that potential $\varphi_{i}\left(\mathbf{k}_{i}, \omega\right)$ is detected in the electron 
oscillating frame as a set of harmonics $\varphi_{i}\left(\mathbf{k}_{i}+q\mathbf{k}_{0}, \omega+q\omega_{0}\right)$.
The IC kinetic parametric instability develops due to the inverse electron Landau damping of these harmonics. 
The frequency $\omega\left(\mathbf{k}_{i}\right)$ of the IC (ion Bernstein) wave is determined as the solution to equation 
\begin{eqnarray}
&\displaystyle 
1+\varepsilon_{i}\left(\mathbf{k}_{i}, \hat{\omega}\right)+\frac{1}{k^2\lambda _{De }^2}=0\label{50} 
\end{eqnarray}
and for $|z_{in}|\gg 1$ is equal approximately to $\hat{\omega}=\omega+C_{i}=\omega\left(\mathbf{k}_{i}\right)=n_{0}\omega_{ci}+\delta\omega
\left(\mathbf{k}_{i}\right)$ with
\begin{eqnarray}
&\displaystyle 
\delta\omega\left(\mathbf{k}_{i}\right)\approx n_{0}\omega_{ci}\frac{I_{n_{0}}\left(k^{2}_{\bot}\rho^{2}_{i}\right)e^{-k^{2}_{\bot}\rho^{2}_{i}}}{\left(1+
\frac{T_{i}}{T_{e}}+k^{2}_{i}\lambda^{2}_{Di}\right)}.\label{51} 
\end{eqnarray}
The growth rate $\gamma\left(\mathbf{k}_{i}\right)$ of the instability is equal approximately to 
\cite{Mikhailenko4}
\begin{eqnarray}
&\displaystyle 
\gamma\left(\mathbf{k}_{i}\right)
\approx 
-\sqrt{\pi}n_{0}\omega_{ci}\frac{T_{i}}{T_{e}}\frac{I_{n_{0}}\left(k^{2}_{\bot}\rho^{2}_{i}\right)e^{-k^{2}_{\bot}
\rho^{2}_{i}}}{\left(1+k^{2}_{i}\lambda^{2}_{Di}+\frac{T_{i}}{T_{e}}\right)^{2}}
\nonumber
\\ 
&\displaystyle
\times
\sum\limits ^{\infty}_{m=-\infty}J^{2}_{m}\left(a_{ei}\right)
\left(z_{e}- m\zeta\right)e^{-\left(z_{e}- m\zeta\right)^{2}}.
\label{52}
\end{eqnarray}
where $z_{e}=\omega\left(\mathbf{k}_{i}\right)/\sqrt{2}k_{iz}v_{Ti}$. As it follows from Eq. (\ref{52}), the growth rate maximum attains 
for $a\sim k_{i\bot}\xi\sim k_{i\bot}U/\omega_{0}\sim k_{i\bot}U/\omega_{ci}
=k_{\bot}\rho_{i}U/v_{ti}\sim 1$, where $U\sim \left| V_{i}-V_{e}\right| $, i. e. for the IC waves with wavelength 
across the magnetic field comparable with the amplitude of the displacements of ions relative to electrons. For the frequency of the FW
$\omega_{0}\sim \omega_{ci}$ these waves are short with $k_{\bot}\rho_{i}>1$ when $v_{Ti}>U$. 
Now, we can present the solution for the potential $\varphi\left(\mathbf{r}_{i},t \right)$, determined by Eq. (\ref{26}), for 
the IC kinetic 
parametric instability, which accounts for the scattering of ions by the IC parametric turbulence,
\begin{eqnarray}
& \displaystyle 
\varphi\left(\mathbf{r}_{i},t \right)=\sum_{n=-\infty}^{\infty}\int
d\mathbf{k}\varphi\left(\mathbf{k}_{i}\right)J_{n}\left(
\frac{k_{i\bot}\bar{v}_{i\bot}}{\omega_{ci}}\right)
\nonumber
\\
&\displaystyle 
\times
\exp\Big[-i\omega\left(\mathbf{k}_{i}\right)t+\gamma t-C_{i}t+ik_{ix}\bar{X}_{i} 
\nonumber
\\ 
&\displaystyle
+ik_{iy}\bar{Y}_{i}+ik_{iz}\left(\bar{z}_{i} +\bar{v}_{iz}t\right)-in
\left(\bar{\phi}-\omega_{ci}t-\theta \right)\Big],
\label{53}
\end{eqnarray}
where $C_{i}$ is determined by Eq. (\ref{31}), which for $|z_{in}|\gg 1$  is equal to
\begin{eqnarray}
&\displaystyle
C_{i}\approx \frac{e^{2}}{2m^{2}_{i}}\sum\limits_{n_{1}=-\infty}^{\infty}\int d\mathbf{k}
_{1}|\varphi\left(\mathbf{k}_{1}\right)|^{2}
\nonumber
\\
&\displaystyle 
\times
\mathcal{F}_{i}\left( k_{\bot},k_{1\bot}\right)e^{-k_{\bot}^{2}\rho^{2}_{i}}\frac{\gamma
\left(\mathbf{k}_{1}\right)}{\left(\delta\omega
\left(\mathbf{k}_{1}\right)\right)^{2}}.
\label{54}
\end{eqnarray}
Equation (\ref{53}) reveals that potential $\varphi\left(\mathbf{r}_{i},t \right)$ ceases its 
growth when $\gamma\left(\mathbf{k}\right)=C_{i}$. For the  conditions of the FW heating in 
the SOL plasma where $v_{ti}/U\sim k_{\bot}\rho_{i}>1$,  
Eq. (\ref{53}) predicts that the energy density $W=\int W\left(\mathbf{k}\right)d\mathbf{k}$ 
of the ion cyclotron turbulence in the 
saturated state, where
\begin{eqnarray}
&\displaystyle
W\left(\mathbf{k}\right)=k^{2}\omega\left(\mathbf{k}\right)
\left| \varphi\left(\mathbf{k}\right)\right| ^{2}
\frac{\partial \varepsilon_{i}}
{\partial \omega\left(\mathbf{k}\right)},
\label{55}
\end{eqnarray}
is given by the estimate\cite{Mikhailenko4}
\begin{eqnarray}
&\displaystyle
\frac{W}{n_{0i}T_{i}}\sim \left(\frac{U}{v_{Ti}}\right)^{4}.
\label{56}
\end{eqnarray}
The development of this instability  results in the heating of the cold SOL ions with heating rate\cite{Mikhailenko1, Mikhailenko4}
\begin{eqnarray}
&\displaystyle n_{0i}\frac{\partial T_{i\bot}}{\partial t}\approx \gamma \frac{W}{n_{0i}
T_{i\bot}}n_{0i}T_{i\bot},
\label{57}
\end{eqnarray} 
resulted from the scattering of ions by the IC turbulence powered by the IC kinetic 
parametric instability.

\subsection{IC quasimode decay instability.}\label{sec3.3}

The main goal of this paper is to elucidate the effect of the IC quasimode decay 
instability\cite{Porkolab1, Porkolab2} in the anomalous absorption of the 
FW energy in the SOL. The development of this instability  is considered 
\cite{Pace,Wilson,Fujii,Rost} as the main mechanism of the transmission of the FW energy to 
the SOL plasma and particularly to the SOL ions.
The dispersion properties of this instability is found 
by the solution of the dispersion equation $\hat{\varepsilon}
\left(\mathbf{k}_{i}, \omega\right)=0$, where $\hat{\varepsilon}\left(\mathbf{k}_{i}, \omega
\right)$ is determined by Eq. (\ref{48}). The equation similar to Eq. (\ref{48}) was derived for the first 
time in Ref.\cite{Porkolab1, Porkolab2} using the equation for the Fourier transform 
$\varphi\left(\mathbf{k}, \omega\right)$ of the potential $\varphi\left(\mathbf{\hat{r}}, t\right)$ determined in the laboratory frame, 
\begin{eqnarray}
&\displaystyle 
\varphi\left(\mathbf{k},\omega\right)+ \sum_{\alpha=i,e}
\sum\limits_{m=-\infty}^{\infty}\sum\limits_{p=-\infty}^{\infty}J_{m}
\left(a_{\alpha}\right)J_{m-p}\left(a_{\alpha}\right)e^{ip\delta_{\alpha}}
\nonumber
\\  
&\displaystyle 
\times
\varepsilon_{\alpha}\left(\mathbf{k}, \omega+m\omega_{0}\right)\varphi \left(\mathbf{k}, 
\omega+p\omega_{0}\right) =0,\label{58}
\end{eqnarray}
Equation  (\ref{58}) contains two parameters $a_{\alpha}$ ($\alpha= i,e$) which are the arguments of the Bessel functions
$J_{m, m-p}\left(a_{\alpha}\right)$,
\begin{eqnarray}
& \displaystyle 
a_{\alpha}=\left\lbrace\left[\frac{e_{\alpha}k_{y}}
{2m_{\alpha}\omega_{0}}\left(\frac{E_{0x}-E_{0y}}
{\omega_{0}-\omega_{c\alpha}} -\frac{E_{0x}+E_{0y}}
{\omega_{0}+\omega_{c\alpha}}\right)\right]^{2}\right. 
\nonumber
\\ 
&\displaystyle
\left.+\left[\frac{e_{\alpha}k_{x}}{2m_{\alpha}\omega_{0}}\left(\frac{E_{0x}+E_{0y}}
{\omega_{0}+\omega_{c\alpha}}+\frac{E_{0x}-E_{0y}}{\omega_{0}-\omega_{c\alpha}}\right) \right]^{2}\right\rbrace ^{1/2},
\label{59}
\end{eqnarray}
and
\begin{eqnarray}
& \displaystyle 
\tan \delta_{\alpha}=\frac{k_{y}{\left(\frac{E_{0x}+E_{0y}}{\omega_{0}+\omega_{c\alpha}}
+\frac{E_{0x}-E_{0y}}{\omega_{0}-\omega_{c\alpha}}\right)}}
{k_{x}\left(\frac{E_{0x}+E_{0y}}{\omega_{0}+\omega_{c\alpha}} 
-\frac{E_{0x}-E_{0y}}{\omega_{0}-\omega_{c\alpha}}\right)}.
\label{60}
\end{eqnarray}
It is contrary to Eq. (\ref{41}), which contains one parameter $a_{ei}$ which determines the 
amplitude of the relative displacements of electron relative to ions in FW.  The parameter 
$a_{\alpha}\sim k\delta r_{\alpha}$, where $\delta r_{\alpha}$ determines the 
amplitude of the displacement of the particle of species $\alpha$ in FW field relative to 
the {\textit{laboratory}} frame. It may be anticipated that the effect of the FW on the 
development of the parametric instabilities is negligible when $a_{\alpha}\ll 1$ for all 
plasma species, and this effect will be strongest when $a_{\alpha}$  is commensurate with 
unity. In last case, all terms in the summations over $m$ and $p$ should be retained, leaving 
undefined what is the solution for the potential $\varphi\left(\mathbf{k}, \omega\right)$. 

The theory of the parametric instabilities, which grounds on Eq. (\ref{58}), 
was developed in Refs. \cite{Porkolab1, Porkolab2}
only for a small values of parameter $a_{\alpha}\ll 1$. This limit, in which only the terms 
with $m=0,\pm 1$ and $p=0,\pm 1$ were accounted for and all other terms in the summation 
over $m, p$ in the ion and electron terms were neglected, corresponds to the limit of 
the small values of the growth rate and may be applied, for example, to the derivations 
of the thresholds of the parametric instabilities. Refereeing on that theory, 
the quasimode decay instabilities were considered\cite{Pace} as a potential 
sources of the generation of the high energy ions in SOL during FW injection in tokamaks. 

It is obvious, that the assessment of the importance of any instability in the processes of 
the anomalous absorption of the RF wave energy and anomalous heating of ions may be made
when, at least, the estimates for the maximum growth rate, for the saturation level of the instability, 
and for the anomalous ion/electron heating rate are known. As the first step on this way
we extend the linear theory\cite{Porkolab1, Porkolab2} of the quasimode decay instabilities 
on the general case of arbitrary values of particles displacements in the FW field by employing Eq. (\ref{48}).

As it was argued in Ref.\cite{Porkolab1, Porkolab2}, the quasimode decay instability occurs when $\varepsilon\left(\mathbf{k}_{i}, \omega\right)\neq 0$ for 
given $\mathbf{k}_{i}$ and $\omega$ in Eq. (\ref{47}), i. e. for which the IC kinetic parametric instability determined by the equation $
\varepsilon\left(\mathbf{k}_{i}, \omega\right)=0$ is absent. It is assumed, however, that for these $\mathbf{k}_{i}$ and $\omega$ the equation
\begin{eqnarray}
&\displaystyle \varepsilon\left(\mathbf{k}_{i}+q_{0}\mathbf{k}_{0}, \omega+q_{0}\omega_{0}\right)=0 
\label{61}
\end{eqnarray}
holds for some value of $q=q_{0}$, with the solution $\omega+q_{0}\omega_{0}=\omega\left(\mathbf{k}_{i}+q_{0}\mathbf{k}_{0}\right)$ 
to Eq. (\ref{50}) with $\omega$ replaced by $\omega+q_{0}\omega_{0}$. 
The coefficient $\varepsilon\left(\mathbf{k}_{i}+q_{0}\mathbf{k}_{0}, \omega+q_{0}\omega_{0}\right)$ in Eq. (\ref{48}) is not 
equal to zero, because the dependence $\omega\left(\mathbf{k}_{i}+q_{0}\mathbf{k}_{0}\right)$ is determined now as a solution 
to the whole equation (\ref{48}). Retaining only the term with $q=q_{0}$ in the summation over $q$ 
in Eq. (\ref{48}) we obtain the relation
\begin{eqnarray}
&\displaystyle 
\varepsilon\left(\mathbf{k}_{i}+q_{0}\mathbf{k}_{0}, \omega+q_{0}\omega_{0}\right)
=\frac{1}{\varepsilon\left(\mathbf{k}_{i}, \omega\right)}
\nonumber
\\ 
&\displaystyle
\times
\sum\limits_{m=-\infty}^{\infty}\sum\limits_{m_{1}=-\infty}^{\infty}J_{m}\left(a_{ei}\right)J_{m_{1}}\left(a_{ei}\right)J_{m+q_{0}}\left(a_{ei}
\right)
\nonumber
\\ 
&\displaystyle
\times J_{m_{1}-q_{0}}\left(a_{ei}\right)\varepsilon_{e}\left(\mathbf{k}_{i}-m\mathbf{k}_{0}, \omega-m\omega_{0}\right)
\nonumber
\\ 
&\displaystyle
\times
\varepsilon_{e}\Big(\mathbf{k}_{i}-\left(m_{1}-q_{0}\right)\mathbf{k}_{0}, \omega-\left(m_{1}-q_{0}\right)\omega_{0}\Big).
\label{62}
\end{eqnarray}
Using the expansion of  $\varepsilon\left(\mathbf{k}_{i}+q_{0}\mathbf{k}_{0}, \omega+q_{0}\omega_{0}\right)$ in the vicinity of the solution 
$\omega=\omega\left(\mathbf{k}_{i}+q_{0}\mathbf{k}_{0}
\right)-q_{0}\omega_{0}$ to equation $\text{Re}\,\varepsilon\left(\mathbf{k}_{i}+q_{0}\mathbf{k}_{0}, \omega\left(\mathbf{k}_{i}+q_{0}\mathbf{k}_{0}\right)
\right)=0$,
\begin{eqnarray}
&\displaystyle 
\varepsilon\left(\mathbf{k}_{i}+q_{0}\mathbf{k}_{0}, \omega+q_{0}\omega_{0}\right)
\nonumber
\\ 
&\displaystyle
=i\gamma\left(\mathbf{k}_{i}+q_{0}\mathbf{k}_{0}\right)
\frac{\partial \text{Re}\, \varepsilon
\left(\mathbf{k}_{i}+q_{0}\mathbf{k}_{0}, \omega\left(\mathbf{k}_{i}+q_{0}\mathbf{k}_{0}\right)\right)}{\partial \omega\left(\mathbf{k}_{i}
+q_{0}\mathbf{k}_{0}\right)}
\nonumber
\\ 
&\displaystyle
+i\text{Im} \,\varepsilon\left(\mathbf{k}_{i}+q_{0}\mathbf{k}_{0}, 
\omega\left(\mathbf{k}_{i}+q_{0}\mathbf{k}_{0}\right)\right)
\label{63}
\end{eqnarray}
we obtain from Eq. (\ref{47}) the growth rate $\gamma\left(\mathbf{k}_{i}+q_{0}\mathbf{k}_{0}\right)$ of the quasimode decay instability,
\begin{eqnarray}
&\displaystyle 
\gamma\left(\mathbf{k}_{i}+q_{0}\mathbf{k}_{0}\right)=\left(\frac{\partial \text{Re}\, \varepsilon
\left(\mathbf{k}_{i}+q_{0}\mathbf{k}_{0}, \omega\left(\mathbf{k}_{i}+q_{0}\mathbf{k}_{0}\right)\right)}{\partial \omega\left(\mathbf{k}_{i}
+q_{0}\mathbf{k}_{0}\right)}\right)^{-1}
\nonumber 
\\
&\displaystyle
\times
\Big[ -\text{Im}\,\varepsilon
\left(\mathbf{k}_{i}+q_{0}\mathbf{k}_{0}, \omega\left(\mathbf{k}_{i}+q_{0}\mathbf{k}_{0}\right)\right)
\nonumber
\\ 
&\displaystyle
+\sum\limits_{m=-\infty}^{\infty}\sum\limits_{m_{1}=-\infty}^{\infty}J_{m}\left(a_{ei}\right)J_{m+q_{0}}
\left(a_{ei}\right)
\nonumber 
\\
&\displaystyle
\times
J_{m_{1}}\left(a_{ei}\right)J_{m_{1}-q_{0}}\left(a_{ei}\right)
\label{64}
\\ 
&\displaystyle
\left.
\times
\text{Im}\,\left(\frac{\varepsilon_{e}\left(\mathbf{k}_{i}- m\mathbf{k}_{0}, 
\omega\left(\mathbf{k}_{i}+q_{0}\mathbf{k}_{0}\right)-\left(m+q_{0}\right)\omega_{0}\right)}
{\varepsilon\left(\mathbf{k}_{i}, \omega\left(\mathbf{k}_{i}+
q_{0}\mathbf{k}_{0}\right)-q_{0}\omega_{0}\right)}
\right.\right.
\nonumber
\\ 
&\displaystyle
\left.\left.
\times
\varepsilon_{e}\Big(\mathbf{k}_{i}-\left(m_{1}-q_{0}\right)
\mathbf{k}_{0}, \omega\left(\mathbf{k}_{i}+q_{0}\mathbf{k}_{0}\right)-m_{1}\omega_{0}\Big)\right)\right]. 
\nonumber
\end{eqnarray}
Equation (\ref{64}) which determines the growth rate of the IC quasimode decay instability is 
much more complicate
than the dispersion equation (\ref{49}) for the IC kinetic parametric instability. In the
limiting case of  small arguments of the Bessel functions, this equation reproduces the 
corresponding equation for the growth rate of the quasimode decay instabilities derived in 
Ref.\cite{Porkolab}, that gives 
the growth rate value near the thresholds. For the deriving the maximum values of the growth 
rate, the quantity necessary for the assessment of this instability in the processes of the 
anomalous absorption of the FW and anomalous heating of ions, the numerical solution of Eq. 
(\ref{64}) is necessary. 

\subsection{Numerical analysis of Eq. (\ref{41}).}\label{sec3.4}
The dispersion equation (\ref{49}) for the IC kinetic parametric instability and Eq. 
(\ref{48}) for the IC quasimode decay instability are the simplest approximations of the 
dispersion equation for the basic equation (\ref{41}) which is in fact the infinite system of 
equations for the potential $\varphi_{i}\left(\mathbf{k}_{i}, \omega\right)$ and infinite 
number of harmonics $\varphi_{i}
\left(\mathbf{k}_{i}-q\mathbf{k}_{0}, \omega-q\omega_{0}\right)$ coupled with $\varphi_{i}
\left(\mathbf{k}_{i}, \omega\right)$. By replacing  $\omega$ 
on $\omega-m\omega_{0}$ in Eq. (\ref{41}), where $m$ is an integer,  Eq.  (\ref{41})
is presented in the form of the infinite system
\begin{eqnarray}
\sum\limits_{q=-\infty}^{\infty}a_{mq}\varphi_{i}\left(\mathbf{k}_{i}-q\mathbf{k}_{0},  
\omega-q\omega_{0}\right)=0,
\label{65}
\end{eqnarray}
where $m$ and $q$ are integer numbers and the coefficients $a_{mq}$ are determined by 
relation
\begin{eqnarray}
&\displaystyle
a_{mq}=\delta_{mq}+\left(1+\varepsilon_{i}\left(\mathbf{k}_{i}-m\mathbf{k}_{0}, 
\omega-m\omega_{0}\right)\right)^{-1}
\nonumber
\\ &
\displaystyle
\times\sum\limits_{r=-\infty}^{\infty}e^{i\left(m-q\right)\left(\pi+\delta\right)}
J_{r+m}\left(a\right)J_{r+q}\left(a\right)
\nonumber
\\ &
\displaystyle
\times
\varepsilon_{e}\left(\mathbf{k}_{i}+r\mathbf{k}_{0}, \omega+r\omega_{0}\right).
\label{66}
\end{eqnarray}
The equality to zero of the determinant of this homogeneous system, 
\begin{eqnarray}
&\displaystyle
\text{det}\left\|a_{mq}\right\| =0,
\label{67}
\end{eqnarray}
gives the general dispersion equation for system (\ref{65}) the solution of which $\omega=
\omega\left(\mathbf{k}_{i}\right)$ determines the dispersive properties of the parametric 
instabilities. Note, that dispersion equations (\ref{42}) corresponds to the accounting for 
in system (\ref{65}) only one term with $q=0$, whereas the result (\ref{64}) for the growth 
rate of the quasimode instability follows from Eq. (\ref{67}) for two-modes system with 
$q=0$ and $q=q_{0}$. 

\begin{figure}[!htbp]
\includegraphics[width=0.4\textwidth]{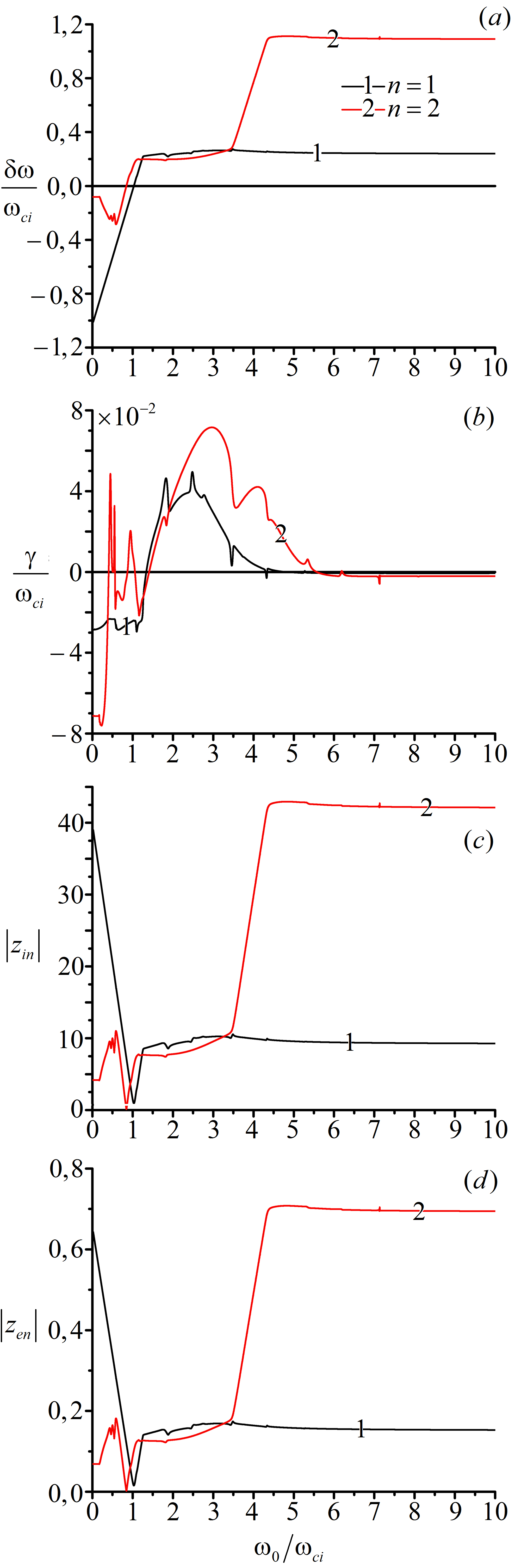}
\caption{\label{fig1} The normalized frequency $\delta\omega/\omega_{ci}$, normalized 
growth rate $\gamma/\omega_{ci}$, $\left|z_{in} \right|$ and $\left|z_{en} \right|$
versus $\omega_{0}/\omega_{ci}$ for $\hat{E}_{0x} = 0.875$ for the first IC 
harmonic, $\hat{E}_{0x} = 0.72$ for second IC harmonic and $\hat{E}_{0y} = 0.3$ for both IC 
harmonics.}
\end{figure}

\begin{figure}[!htbp]
\includegraphics[width=0.4\textwidth]{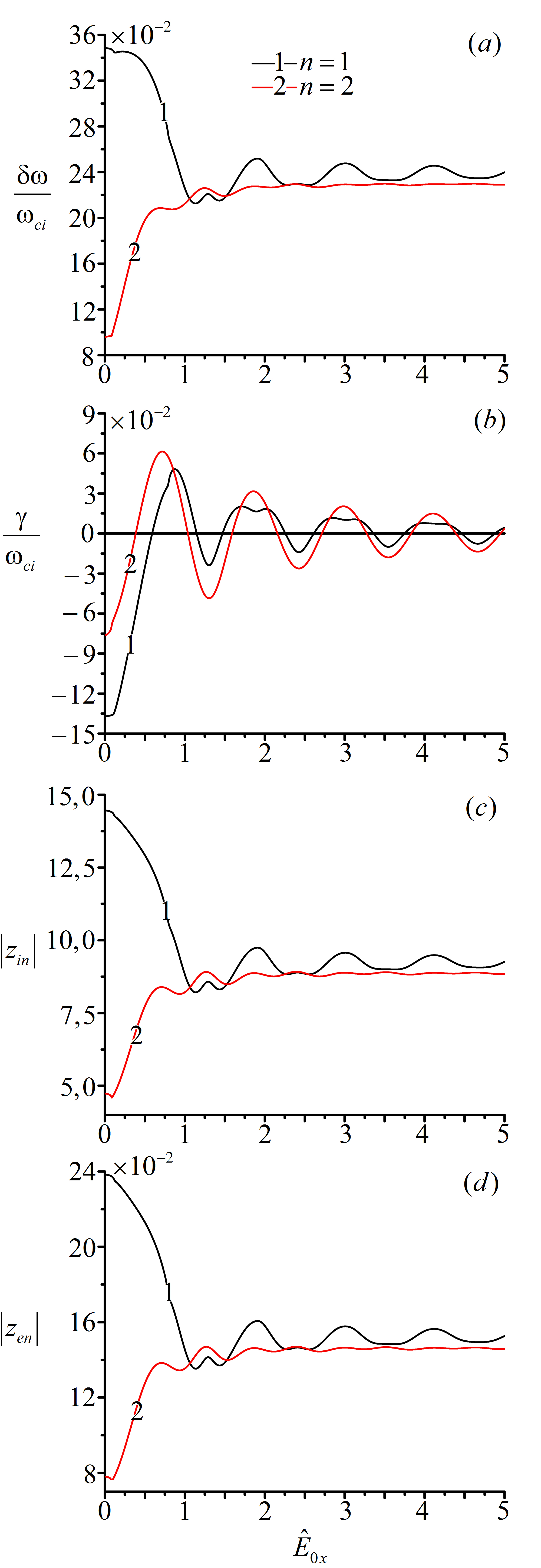}
\caption{\label{fig2} The normalized frequency $\delta\omega/\omega_{ci}$, 
normalized growth rate $\gamma/\omega_{ci}$, $\left|z_{in} \right|$ 
and $\left|z_{en} \right|$ versus $\hat{E}_{0x}$ for $\hat{E}_{0y} = 0.3$ and $\omega_{0}/
\omega_{ci}=2.5$.}
\end{figure}

\begin{figure}[!htbp]
\includegraphics[width=0.4\textwidth]{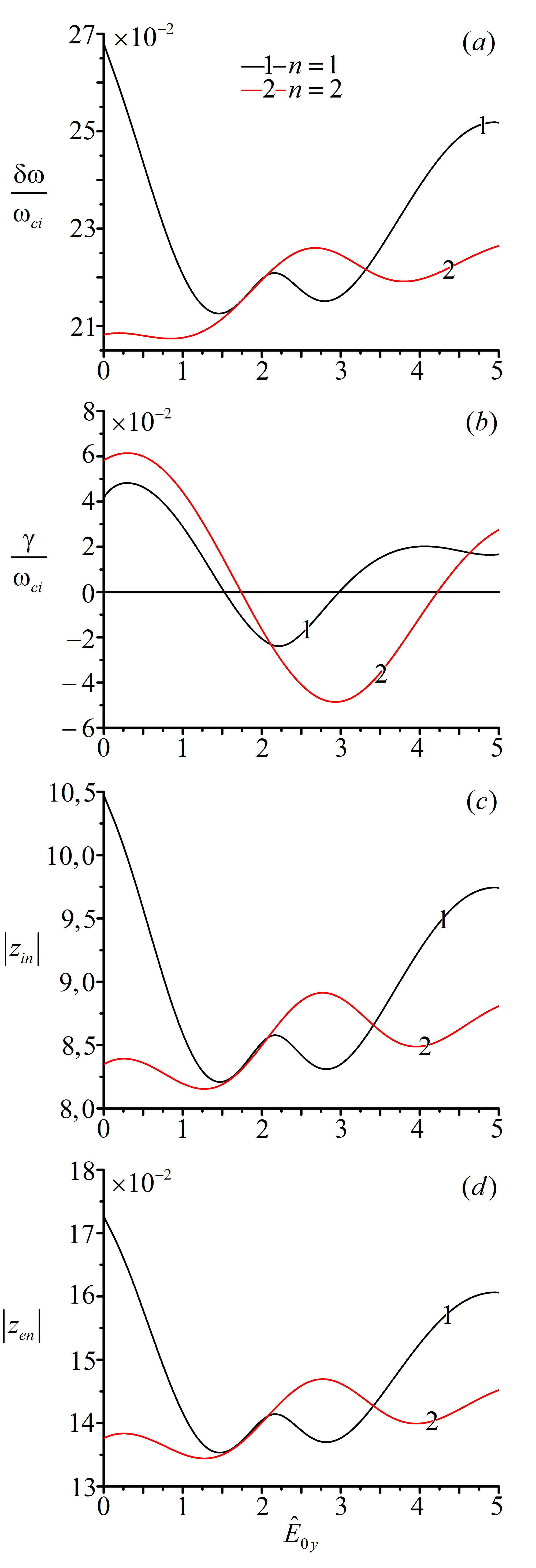}
\caption{\label{fig3} The normalized frequency $\delta\omega/\omega_{ci}$, 
normalized growth rate $\gamma/\omega_{ci}$, $\left|z_{in} \right|$ 
and $\left|z_{en} \right|$	versus $\hat{E}_{0y}$ for $\hat{E}_{0x} = 0.875$ for the first IC 
harmonic, $\hat{E}_{0x} = 0.72$ for second IC harmonic and $\omega_{0}/\omega_{ci}=2.5$.}
\end{figure}

\begin{figure}[!htbp]
\includegraphics[width=0.4\textwidth]{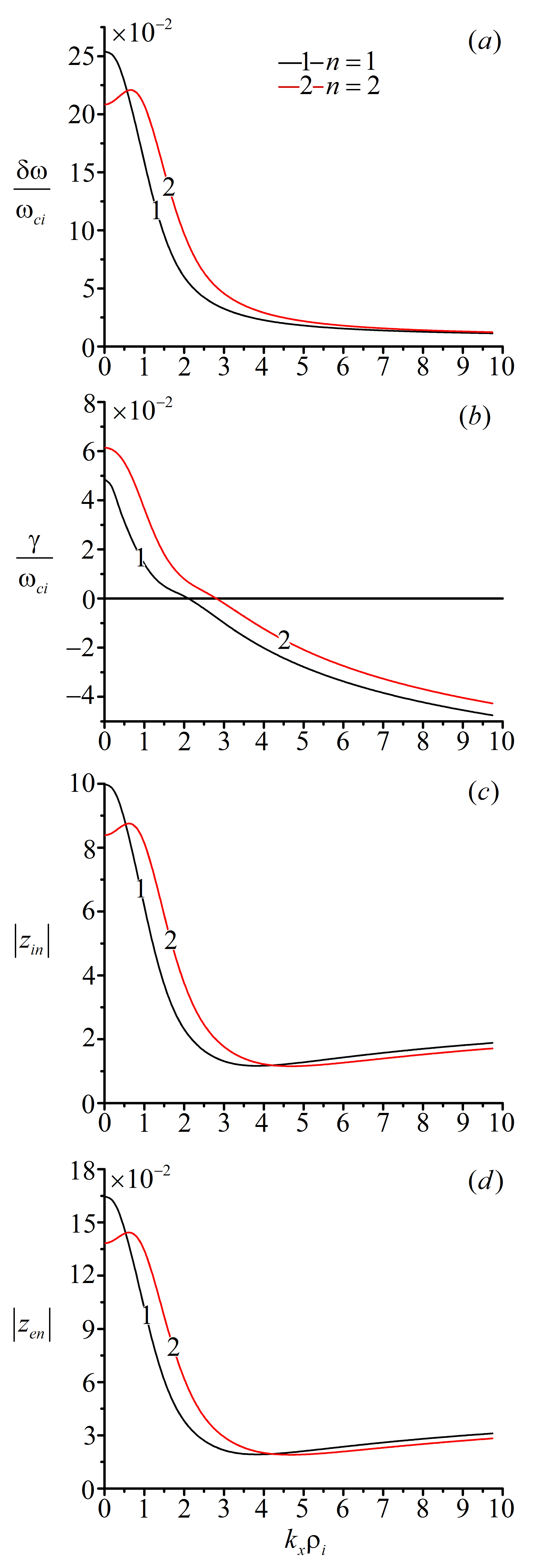}
\caption{\label{fig4} The normalized frequency $\delta\omega/\omega_{ci}$, 
normalized growth rate $\gamma/\omega_{ci}$, $\left|z_{in} \right|$ 
and $\left|z_{en} \right|$ versus $k_{x}\rho_{i}$.}
\end{figure}

\begin{figure}[!htbp]
\includegraphics[width=0.4\textwidth]{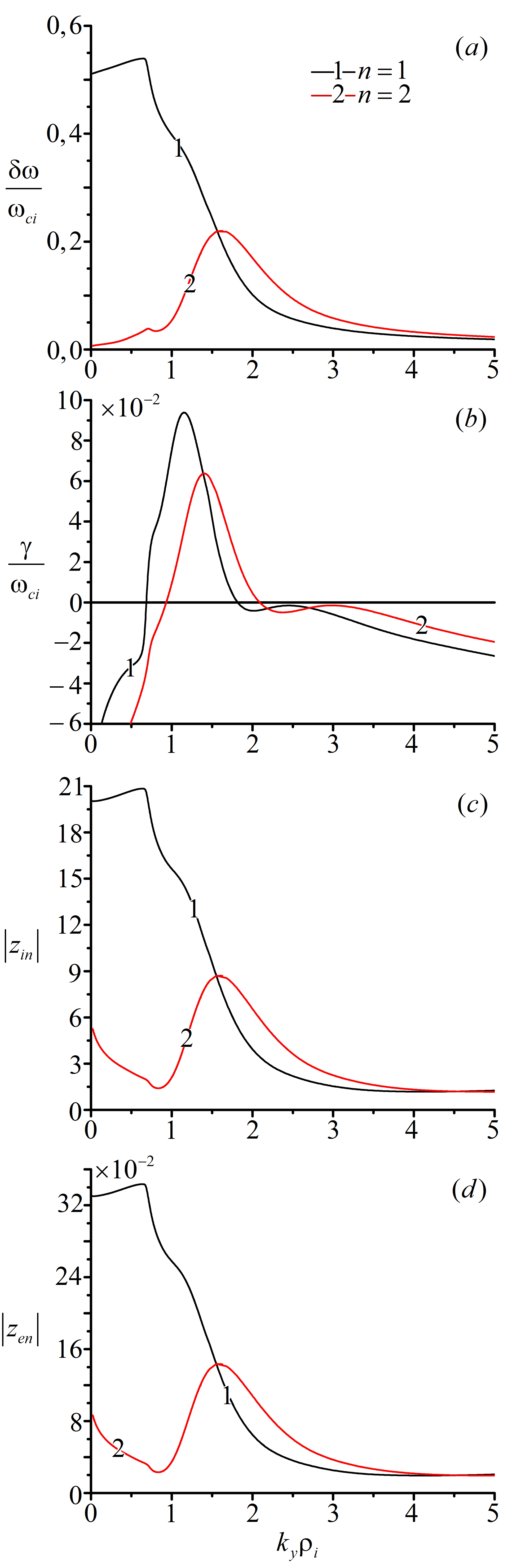}
\caption{\label{fig5} The normalized frequency $\delta\omega/\omega_{ci}$, 
normalized growth rate $\gamma/\omega_{ci}$, $\left|z_{in} \right|$ 
and $\left|z_{en} \right|$ versus $k_{y}\rho_{i}$.}
\end{figure}

\begin{figure}[!htbp]
\includegraphics[width=0.4\textwidth]{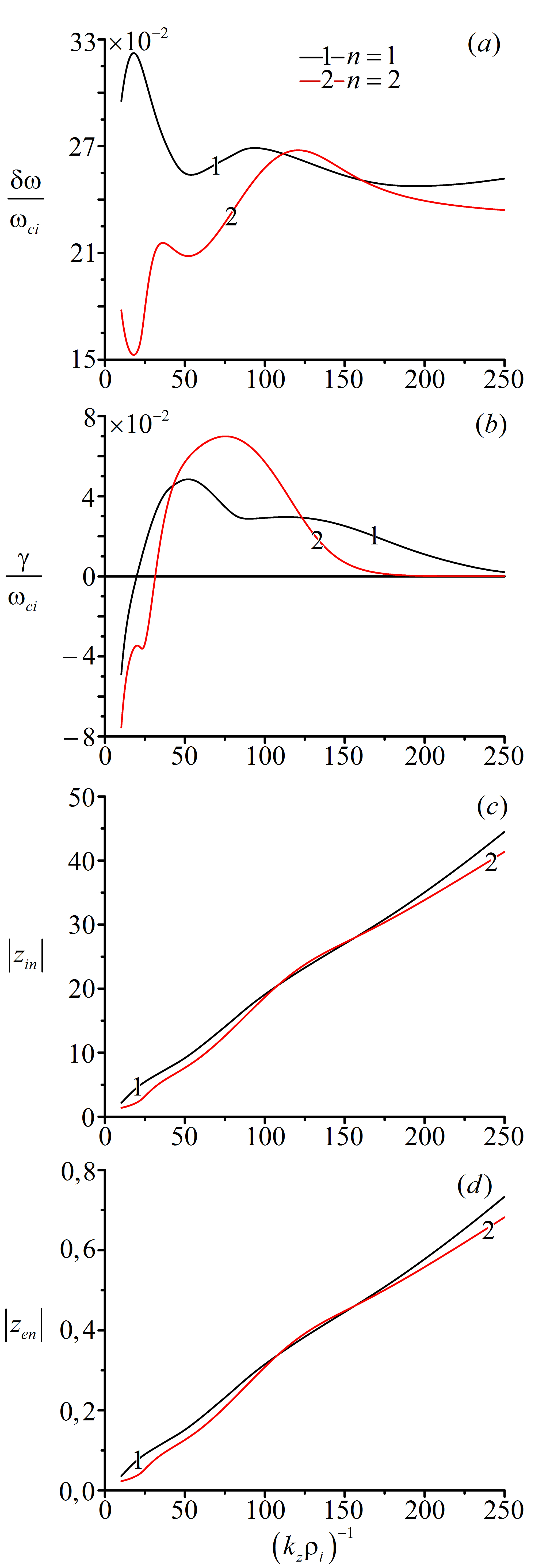}
\caption{\label{fig6} The normalized frequency $\delta\omega/\omega_{ci}$, 
normalized growth rate $\gamma/\omega_{ci}$, 
$\left|z_{in} \right|$ and $\left|z_{en} \right|$
versus $\left( k_{z}\rho_{i}\right) ^{-1}$.}
\end{figure}

\begin{figure}[!htbp]
\includegraphics[width=0.4\textwidth]{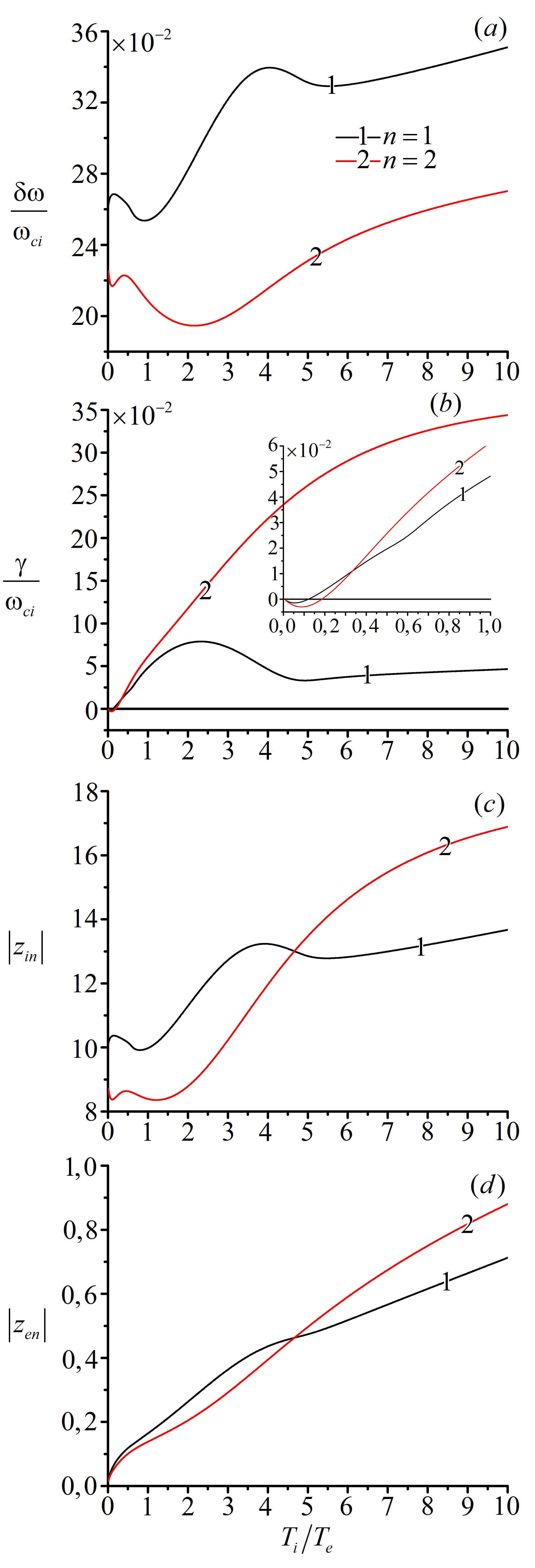}
\caption{\label{fig7} The normalized frequency $\delta\omega/\omega_{ci}$, normalized growth 
rate $\gamma/\omega_{ci}$, $\left|z_{in} \right|$ and $\left|z_{en} \right|$
versus $T_{i}/T_{e}$.}
\end{figure}

In this subsection, we present the results of the numerical solution of 
Eq. (\ref{67}) for the three-modes system which contains the fundamental mode $
\varphi_{i}\left(\mathbf{k}_{i}, \omega\right)$ and harmonics $\varphi_{i}\left(\mathbf{k}
_{i},  \omega-\omega_{0}\right)$, $\varphi_{i}\left(\mathbf{k}_{i},  \omega-2\omega_{0}
\right)$. The results are presented in Figs. \ref{fig1}--
\ref{fig7} for the first two IC harmonics. In all figures, the black line (line 1) denote the 
results for the first IC harmonic $\left(n=1\right)$, and the red line (line 2) denote the 
results for the second IC harmonic $\left(n=2\right)$. In all these figures, the solution for 
the normalized frequency $\delta\omega\left(\mathbf{k}\right)/\omega_{ci}$, where $\delta
\omega\left(\mathbf{k} \right)= {\text{Re}}\,\omega\left(\mathbf{k}\right)-n
\omega_{ci}$ is presented in panel (a), the normalized growth 
rate $\gamma\left(\mathbf{k}\right)/\omega_{ci}$, where $\gamma\left(\mathbf{k}
\right)={\text{Im}}\,\omega\left(\mathbf{k}\right)$, is presented in panel (b), 
and the arguments $|z_{in}|=|\left(\omega\left(\mathbf{k}\right)- 
n\omega_{ci}\right)|/\sqrt{2}k_{z}v_{Ti} $ and  $|z_{en}|=|n\omega_{ci}+\delta\omega
\left(\mathbf{k}_{i}\right)|/\sqrt{2}k_{iz}v_{Ti}$ of the $W$-functions in $
\varepsilon_{i}$ and $\varepsilon_{e}$ are presented in panels (c) and (d) respectively.
In Figs. \ref{fig1}--\ref{fig6} the results are presented for a plasma with equal ion and 
electron temperature, $T_{i}/T_{e}=~1$, ion/electron mass ratio $m_{i}/m_{e} = 2\cdot 1840$, 
SOL ion density $n_{0i} = 2\cdot 10^{10}cm^{-3}$ and the magnitude of the magnetic field 
$B_{0} = 1\,T$. In Fig. \ref{fig1}, the solution to Eq. (\ref{67}) for the normalized 
frequency, the normalized growth rate, and $|z_{in}|$, $|z_{en}|$  versus normalised FW 
frequency $\omega_{0}/\omega_{ci}$ for $k_{x}\rho_{i} = 0.7$, $k_{y}\rho_{i} = 1.47$; 
$\left( k_{z}\rho_{i}\right) ^{-1} = 54.6$ are given. In our calculations we used the 
normalised values of the FW electric fields,
\begin{eqnarray}
&\displaystyle 
\hat{E}_{0x,y}=\frac{E_{0x,y}}{\sqrt{4\pi n_{0i}T_{i}}}.
\label{68}
\end{eqnarray}
In Fig. \ref{fig1}, we use $\hat{E}_{0x} = 0.875$ for the first IC 
harmonic, $\hat{E}_{0x} = 0.72$ for the second IC harmonic and $\hat{E}_{0y} = 0.3$ 
for both IC harmonics. These normalised values correspond 
for plasma with $n_{0i}=10^{10}\,cm^{-3}$ and $T_{i}=20\,eV$ to $E_{0y}=180 V/cm$, and to 
$E_{0x} = 525\,V/cm$ and to $E_{0x} = 432\,V/cm$ respectively.
In the summation over $n$ in $\varepsilon_{i}$ determined by Eq. (\ref{43}), we account for 
all terms in the interval $[-30;30]$, and in the summation over $r$ in Eq. (\ref{66}) we 
account for the terms in the interval $[-50;50]$. Figure \ref{fig1} reveals that both IC 
harmonics are unstable in the finite intervals  $\omega_{0}/\omega_{ci}$ values above and 
below the unity with the growth rate less than $\delta\omega\left(\mathbf{k}\right)$. The 
$\omega_{0}/\omega_{ci}=2.5$ value, where the growth rates of both IC harmonics are almost  
maximum, is used in Figs. \ref{fig2}--\ref{fig7} as the optimal value for the normalized FW 
frequency.

In Fig. \ref{fig2}, the solution to Eq. (\ref{67}) for the normalized 
frequency, the normalized growth rate, and $|z_{in}|$, $|z_{en}|$  versus $E_{0x}$ are given 
for $\hat{E}_{0y} = 0.3$, $\omega_{0}/\omega_{ci}=2.5$, $k_{x}\rho_{i} = 0.7$, $k_{y}\rho_{i} 
= 1.47$ and $\left( k_{z}\rho_{i}\right) ^{-1} = 54.6$. It follows from Fig. \ref{fig2}, that 
the growth rate maximum attains at $\hat{E}_{0x} = 0.875$ for the first IC harmonic and at 
$\hat{E}_{0x} = 0.72$ for the second IC harmonic. These values of $\hat{E}_{0x}$ are employed 
in the calculations presented in Figs. \ref{fig1} and \ref{fig3} -\ref{fig7}. 

The results of Eq. (\ref{67}) solution are plotted in Fig. \ref{fig3} for the normalized 
frequency, for the normalized growth rate, for $|z_{in}|$ and  $|z_{en}|$
versus $\hat{E}_{0y}$ for $\omega_{0}/\omega_{ci}=2.5$,  $k_{x}\rho_{i} = 0.7$, $k_{y}
\rho_{i}= 1.47$ and $\left( k_{z}\rho_{i}\right) ^{-1} = 54.6$. 
In these calculations, we used the values $\hat{E}_{0x} = 0.875$ for the first IC harmonic 
and  $\hat{E}_{0x} = 0.72$ for the second IC harmonic. Figure \ref{fig3} confirms that for 
these values of $\hat{E}_{0x}$ the maximum growth rates for both IC harmonics occurs at $
\hat{E}_{0y}=0.3$. 

In Fig. \ref{fig4}, the solution to Eq. (\ref{67}) for the normalized 
frequency, the normalized growth rate, and $|z_{in}|$, $|z_{en}|$  versus $k_{x}\rho_{i}$
are given for $\omega_{0}/\omega_{ci}=2.5$, $k_{y}\rho_{i} = 1.47$ and $\left( k_{z}\rho_{i}
\right) ^{-1} = 54.6$. It follows from Fig. \ref{fig4}, that the discovered parametric IC 
instability has the maximum growth rate for $k_{x}\rho_{i}\ll 1$ and exists for the used 
plasma and FW parameters at $k_{x}\rho_{i}<2$ for the first IC harmonic and at $k_{x}\rho_{i}
<3$ for the second IC harmonic. The lower value for $k_{x}\rho_{i}$ is limited by the 
thickness of the SOL layer. In our calculations presented in other Figures we used $k_{x}
\rho_{i}=0.7$, that corresponds to $k_{x}\approx 10\,cm^{-1}$ for $B_{0} = 1\,T$ and $T_{i}
=20\,eV$. 

In Fig. \ref{fig5}, the solution to Eq. (\ref{67}) are presented for the normalized 
frequency, the normalized growth rate, for $|z_{in}|$ and  $|z_{e}|$ versus $k_{y}\rho_{i}$ 
for first two IC harmonics. This Figure reveals that the IC instability develops in the 
finite interval $\Delta k_{y}\sim k_{y}$ of the $k_{y}\rho_{i}$ values  with $k_{y}\rho_{i}
\approx 1$ for the maximum growth rate of the first IC harmonic and $k_{y}\rho_{i}=1.47$ for 
the second IC harmonic. Figure \ref{fig5} explains the reason for the $k_{y}\rho_{i}= 1.47$ 
value selection in all Figures excluding Fig. \ref{fig5}. This value of $k_{y}\rho_{i}$ is 
optimal at which the growth rates of both IC modes are in the region of their maximum values.

Figure \ref{fig6} reveals that both IC harmonics of the discovered IC instability are 
unstable in the limited interval of the $\left(k_{z}\rho_{i}\right)^{-1}$ values. It explains 
why  $\left(k_{z}\rho_{i}\right)^{-1}= 54.6$ value was selected in all Figures excluding Fig. 
\ref{fig6}.

Figure \ref{fig7} displays that the discovered IC instability is absent in plasma with cold 
ions with $T_{i}/T_{e}< 0.2$. The growth rate for the first IC harmonic has maximum growth 
rate for $T_{i}/T_{e}\sim 2$ and the growth rate for the second IC harmonic  attains its 
maximum at larger $T_{i}/T_{e}$ ratio.

The common conclusion, which follows from all presented Figures, may be given for the 
magnitudes of the parameters $|z_{in}|$ and $|z_{en}|$. We found that for the discovered 
instability $|z_{in}|\gg 1$ and $|z_{en}|<1$. It means that the inverse electron Landau 
damping is decisive process in the 
development of this instability and the IC damping of the unstable IC waves is negligible 
small. In this case, the renormalized version (\ref{43}) for $\varepsilon_{i}$ which accounts 
for the effect of the ion scattering by the ensemble of the IC waves should be employed in 
Eq. (\ref{67}). Because for the maximum growth rate of this instability $k_{\bot}\rho_{i}
\approx k_{y}\rho_{i}\approx  1.5$, the equation $\gamma = C_{i}$ with the growth rate for 
the most unstable IC waves and $C_{i}$ given by Eq. (\ref{54}) gives the estimate 
\begin{eqnarray}
&\displaystyle
\frac{W}{n_{0i}T_{i}}\sim 1
\label{69}
\end{eqnarray}
for the energy density in the saturation state of the IC turbulence powered by this IC 
instability


\section{Turbulent heating of ions by IC parametric turbulence powered by the IC quasimode decay instability.}\label{sec4}

The eventual purpose of our analysis of the IC parametric instabilities is to understand the role of these instabilities in the 
turbulent heating of ions in SOL and generation of the population of the high 
energy ions usually observed in the SOL during FW heating.

In was found in Sec \ref{sec3}, that the Vlasov equation in the ion oscillating frame has a 
form as for a plasma without FW. Therefore the equation of the quasilinear theory for 
the equilibrium ion distribution function $F_{0i}$, which accounts for the effects of the ion 
cyclotron turbulence has a known form 
\begin{eqnarray}
&\displaystyle 
\frac{\partial F_{i0}}{\partial t}+\frac{e}{m_{i}}\frac{\omega_{ci}}{v_{\bot}}\left\langle 
\frac{\partial\varphi}{\partial\phi}
\frac{\partial f_{i}}{\partial v_{\bot}}-\frac{\partial\varphi}{\partial v_{\bot}}\frac{\partial f_{i}}{\partial \phi}\right\rangle
\nonumber
\\ 
&\displaystyle
-\frac{e}{m_{i}}\left\langle\frac{\partial\varphi}{\partial z_{i}}\frac{\partial f_{i}}{\partial v_{iz}}\right\rangle=0,
\label{70}
\end{eqnarray}
where $f_{i}\left(v_{\bot}, \phi, v_{z}, X_{i}, Y_{i}, z_{i}, t\right)$ is 
determined by the equation 
\begin{eqnarray}
&\displaystyle 
f_{i}\left(v_{\bot}, \phi, v_{z}, X_{i}, Y_{i}, z_{i}, t\right)
=\frac{ie}{m_{i}}\sum\limits_{n=-\infty}^{\infty}
\nonumber
\\ 
&\displaystyle
\times
\int\limits_{t=t_{0}}^{t}
dt_{1}
\int d\mathbf{k}_{i}\int d\omega J_{n}\left(\frac{k_{i\bot}v_{\bot}}{\omega_{ci}}\right)
\varphi
\left(\mathbf{k}_{i}, \omega\right)
\nonumber
\\ 
&\displaystyle
\times
e^{i\Psi}
\left(\frac{n\omega_{ci}}{v_{\bot}}\frac{\partial F_{i0}}{\partial v_{\bot}}+k_{iz}
\frac{\partial F_{i0}}{\partial v_{z}}\right)
\label{71}
\end{eqnarray}
where 
\begin{eqnarray}
&\displaystyle 
\Psi= -\left(\omega - k_{iz}v_{z}-n\omega_{ci}\right)t_{1}
-n\left(\phi_{1}-\theta_{1}\right)
\nonumber
\\ 
&\displaystyle
+k_{ix}X_{i}+k_{iy}Y_{i}+k_{iz}z_{1}.
\nonumber
\end{eqnarray}

The quasilinear equation for the $F_{0i}$ for the IC turbulence powered by the kinetic 
parametric IC instability was obtained in Refs. \cite{Mikhailenko1}. Here, we derive the 
quasilinear equation for $F_{0i}$ for the case when IC turbulence is powered by the IC 
quasimode decay instability. The potential $\varphi\left(\mathbf{k}_{i},\omega\right)$
in Eq. (\ref{71}) is coupled with the quasimode potential 
\begin{eqnarray}
&\displaystyle 
\varphi\left(\mathbf{K}, \Omega\right)=\varphi\left(\mathbf{K}\right)
\delta\left(\Omega-\omega\left(\mathbf{K}\right)\right), 
\label{72}
\end{eqnarray}
$\mathbf{K}=\mathbf{k}_{i}+q_{0}\mathbf{k}_{0}$ and $\Omega=\omega+q_{0}\omega_{0}$, by Eq. 
(\ref{41}),
\begin{eqnarray}
&\displaystyle 
\varphi_{i}\left(\mathbf{k}_{i}, \omega\right)
=-\frac{1}{\varepsilon\left(\mathbf{k}_{i}, \omega\right)}\sum\limits_{m=-\infty}^{\infty}
J_{m}\left(a_{ei}\right)J_{m+q_{0}}\left(a_{ei}\right)
\nonumber
\\ 
&\displaystyle
\times 
\varepsilon_{e}\Big(\mathbf{K}-\left(m+q_{0}\right)\mathbf{k}_{0}, \Omega-\left(m+q_{0}\right)\omega_{0}\Big)
\nonumber
\\ 
&\displaystyle
\times
e^{iq_{0}\delta}
\delta\left(\Omega-\omega\left(\mathbf{K}\right)\right)\varphi\left(\mathbf{K}\right),
\label{73}
\end{eqnarray}
in which only the term with $q=q_{0}$ is accounted for in the summation over $q$. The 
essential element in the derivation of Eq. (\ref{70}) with $f_{i}$
and $\varphi$ determined by Eqs. (\ref{71}) -- (\ref{73}), is the calculation of the 
correlation function $\left\langle \varphi_{i}\left(\mathbf{k}_{i1}, 
\omega_{1}\right)\varphi_{i}\left(\mathbf{k}_{i2}, \omega_{2}\right) \right\rangle$. Using 
Eq. (\ref{73}) we obtain that
\begin{eqnarray}
&\displaystyle 
\left\langle \varphi_{i}\left(\mathbf{k}_{i1}, \omega_{1}\right)\varphi_{i}\left(\mathbf{k}
_{i2}, \omega_{12}\right)\right\rangle
\nonumber
\\ 
&\displaystyle
=-\frac{1}
{\varepsilon\left(\mathbf{K}_{1}-q_{0}\mathbf{k}_{0}, \omega\left(\mathbf{K}_{1}\right)-q_{0}
\omega_{0}\right)}
\nonumber
\\ 
&\displaystyle	
\times\frac{1}{\varepsilon\left(\mathbf{K}_{2}-q_{0}
\mathbf{k}_{0}, \omega\left(\mathbf{K}_{2}\right)-q_{0}\omega_{0}\right)}\nonumber
\\ 
&\displaystyle
\times\sum\limits_{m_{1}=-\infty}^{\infty}\sum\limits_{m_{2}=-\infty}^{\infty}
J_{m_{1}}\left(a_{ei}\right)J_{m_{1}+q_{0}}\left(a_{ei}\right)
\nonumber
\\ 
&\displaystyle
\times
J_{m_{2}}\left(a_{ei}\right)J_{m_{2}-q_{0}}\left(a_{ei}\right)
\nonumber
\\ 
&\displaystyle
\times
\varepsilon_{e}\Big(\mathbf{K}_{1}-\left(m_{1}+q_{0}\right)\mathbf{k}_{0}, \omega\left(\mathbf{K}_{1}\right)
-\left(m_{1}+q_{0}\right)\omega_{0}\Big)
\nonumber
\\ 
&\displaystyle
\times\varepsilon_{e}\Big(\mathbf{K}_{2}-\left(m_{2}+q_{0}\right)\mathbf{k}_{0}, \omega\left(\mathbf{K}_{2}\right)
-\left(m_{2}+q_{0}\right)\omega_{0}\Big)
\nonumber
\\ 
&\displaystyle
\times
\left| \varphi\left(\mathbf{K}_{1}\right)\right| ^{2}
\delta\left(\Omega_{1}-\omega\left(\mathbf{K}_{1}\right)\right)\delta\left(\Omega_{2}-\omega\left(\mathbf{K}_{2}
\right)\right)
\nonumber
\\ 
&\displaystyle
\times
\delta\left(\mathbf{K}_{1}+\mathbf{K}_{2}\right)\delta\left(\Omega_{1}+\Omega_{2}\right).
\label{74}
\end{eqnarray}
With correlation function (\ref{74}), the quasilinear equation (\ref{70}) for the ion distribution function $F_{0i}$ becomes
\begin{eqnarray}
&\displaystyle 
\frac{\partial F_{i0}}{\partial t}=\frac{\pi e^{2}}{m^{2}_{i}}\int d\mathbf{k}\sum\limits_{m_{1}=-\infty}^{\infty}\sum\limits_{m_{2}=-
\infty}^{\infty}\sum\limits_{n=-\infty}^{\infty}
\nonumber 
\\
&\displaystyle
\times
J_{m_{1}}\left(a_{ei}\right)J_{m_{1}+q_{0}}\left(a_{ei}\right)
J_{m_{2}}\left(a_{ei}\right)J_{m_{2}-q_{0}}\left(a_{ei}\right)
\nonumber
\\ 
&\displaystyle
\times
\left| \varphi\left(\mathbf{k}_{1}+q_{0}\mathbf{k}_{0}\right)\right| ^{2}
\nonumber 
\\
&\displaystyle
\times
\frac{\varepsilon_{e}\Big(\mathbf{k}-m_{1}\mathbf{k}_{0}, \omega\left(\mathbf{k}+q_{0}
\mathbf{k}_{0}\right)
-\left(m_{1}+q_{0}\right)\omega_{0}\Big)
}{\left| \varepsilon\Big(\mathbf{k}, \omega\left(\mathbf{k}+q_{0}\mathbf{k}_{0}\right)-q_{0}\omega_{0}\Big)\right| ^{2}}
\nonumber
\\ 
&\displaystyle
\times\varepsilon_{e}\Big(-\mathbf{k}+m_{2}\mathbf{k}_{0}, -\omega\left(\mathbf{k}+q_{0}\mathbf{k}_{0}\right)
-\left(m_{1}-q_{0}\right)\omega_{0}\Big)
\nonumber
\\ 
&\displaystyle
\times\left(\frac{n\omega_{ci}}{v_{\bot}}\frac{\partial}{\partial v_{\bot}}+k_{iz}\frac{\partial}{\partial v_{z}}\right)
\nonumber 
\\
&\displaystyle
\times
J^{2}_{n}\left(\frac{k_{\bot}
v_{\bot}}{\omega_{ci}}\right)\left(\frac{n\omega_{ci}}{v_{\bot}}\frac{\partial F_{i0}}
{\partial v_{\bot}}+k_{iz}\frac{\partial F_{i0}}{\partial v_{z}}
\right)
\nonumber
\\ 
&\displaystyle
\times \delta\Big(\omega\left(\mathbf{k}+q_{0}\mathbf{k}_{0}\right)-q_{0}\omega_{0}-n
\omega_{ci}-k_{z}v_{z}\Big),
\label{75}
\end{eqnarray}
where the relation 
\begin{eqnarray}
&\displaystyle 
\left(\omega\left(\mathbf{k}+q_{0}\mathbf{k}_{0}\right)-q_{0}\omega_{0}-n\omega_{ci}-k_{z}v_{z}\right)^{-1}
\nonumber
\\ 
&\displaystyle
=\frac{P}{\left(\omega
\left(\mathbf{k}+q_{0}\mathbf{k}_{0}\right)-q_{0}\omega_{0}-n\omega_{ci}-k_{z}v_{z}\right)}
\nonumber
\\ 
&\displaystyle
-i\pi\delta\left(\omega\left(\mathbf{k}+q_{0}\mathbf{k}_{0}
\right)-q_{0}\omega_{0}-n\omega_{ci}-k_{z}v_{z}\right)
\label{76}
\end{eqnarray}
was used. The quasilinear equation (\ref{75}) is the basic equation for the
derivation, as the corresponding moments of that equation,
the equations which govern the temporal evolution of the ion thermal energy. By
multiplying Eq. (75) on $m_{i}v^{2}_{i\bot}/2$ and integrating this equation over velocities $\mathbf{v}_{i}$, we obtain the
equation 
\begin{eqnarray}
&\displaystyle n_{0i}\frac{\partial T_{i\bot}}{\partial t}\approx \int d\mathbf{k}\gamma\left(\mathbf{k}+q_{0}\mathbf{k}_{0}\right)W\left(\mathbf{k}+q_{0}\mathbf{k}_{0}\right)
\frac{k^{2}}{\left(\mathbf{k}+q_{0}\mathbf{k}_{0}\right)^{2}}
\nonumber
\\ 
&\displaystyle
\times
\frac{\left(q_{0}\omega_{0}-\omega\left(\mathbf{k}+q_{0}\mathbf{k}_{0}
\right)\right)}{\omega\left(\mathbf{k}+q_{0}\mathbf{k}_{0}\right)}\sim \gamma \frac{W}{n_{0i}T_{i\bot}}n_{0i}T_{i\bot},
\label{77}
\end{eqnarray}
which determines the temporal evolution of the ion temperature $T_{i\bot}$ across the magnetic field. 
This estimate was obtained with assumption that the 
spectrum of fluctuations is peaked near the linearly most unstable
wavenumber. The equation which determines 
the temporal evolution of the ion temperature $T_{iz}$ along the magnetic field 
\begin{eqnarray}
&\displaystyle n_{0i}\frac{\partial T_{iz}}{\partial t}\approx \int d\mathbf{k}\gamma\left(\mathbf{k}+q_{0}\mathbf{k}_{0}
\right)\frac{k^{2}}{\left(\mathbf{k}+q_{0}\mathbf{k}_{0}\right)^{2}}
\nonumber
\\ 
&\displaystyle
\times
\frac{\left(q_{0}\omega_{0}+n\omega_{ci}-\omega\left(\mathbf{k}+q_{0}\mathbf{k}_{0}
\right)\right)}{\omega\left(\mathbf{k}+q_{0}\mathbf{k}_{0}\right)}W\left(\mathbf{k}+q_{0}
\mathbf{k}_{0}\right)
\nonumber
\\ 
&\displaystyle
\sim k_{z}\rho_{i}\gamma W\ll \gamma 
\frac{W}{n_{0i}T_{i\bot}}n_{0i}T_{i\bot},
\label{78}
\end{eqnarray}
is derived by the multiplying Eq. (\ref{74}) on $m_{i}v^{2}_{z}/2$ and integrating over the 
velocities $\mathbf{v}_{i}$. 

Eqs. (\ref{77}) and (\ref{78}) reveal that the edge plasma ions 
are heated by the IC parametric turbulence powered by 
the IC quasimode decay instability and this heating is anisotropic with $T_{i\bot}\gg T_{i
z}$. Similar result was derived earlier in Refs\cite{Mikhailenko1, Mikhailenko4} for the IC 
parametric turbulence powered by the kinetic parametric instability. Equations (\ref{57}) 
(\ref{77}) and (\ref{78}) give the order of value estimates for the maximum values of the 
heating rates of SOL ions by the IC parametric turbulence.

\section{Conclusions}\label{sec5}
In this paper, we consider the effect of the IC parametric instabilities on the anomalous 
heating of ions in SOL and on the generation of the bursts of suprathermal ions.

Using the methodology of the oscillating modes we develop the theory of the IC parametric 
instabilities for the general case of large displacements of the electrons relative to ions 
commensurable with the wavelength of the unstable IC waves. As the application of this 
theory we found the expression (\ref{64}) for the growth rate for the IC quasimode decay 
instability which is not limited by the requirement of the small displacements of particles 
in the FW. 

The numerical solution of the general dispersion equation (\ref{67}) for the three IC waves 
system which contains the fundamental mode $\varphi_{i}\left(\mathbf{k}_{i}, \omega\right)$ 
and harmonics $\varphi_{i}\left(\mathbf{k}_{i},  \omega-\omega_{0}\right)$, $\varphi_{i}
\left(\mathbf{k}_{i}, \omega-2\omega_{0}\right)$ reveals that the maximum growth rate has 
the short wavelength  IC instability with $k_{i\bot}\rho_{i}$ of the order of unity. The
inverse electron Landau damping plays essential role in the development of this instability.
The possible mechanism of the saturation of this instability is the scattering of ions by the 
ensemble of the IC waves with random phases, which limits the development of the instability 
on the high level (\ref{69}). 

The estimates for the anomalous  heating rates of ions in SOL are derived using the developed  
quasilinear theory for the ion distribution function which evolves under the action of the IC 
turbulence powered by the quasimode decay instability. The estimates (\ref{77}), (\ref{78}) 
for the maximum anomalous heating rate of ions in SOL confirm the observation of the strong 
anisotropic heating of the cold ions in SOL\cite{Wilson}. At the same time, this heating is 
negligible small for to be responsible for the generation of the bursts of the suprathermal 
ions.

\begin{acknowledgments}
This work was supported by National R\&D Program through the National Research Foundation of 
Korea (NRF) funded by the Ministry of Education, Science and Technology (Grant No. 
NRF--2018R1D1A3B07051247).
\end{acknowledgments}

{}

\begin{thebibliography}{}

\bibitem{Perkins1} R.~J.~Perkins, J.~C.~Hosea, G.~J.~Kramer, J.~-W.~Ahn, R.~E.~Bell, A.~Diallo, S.~Gerhardt,
T.~K.~Gray, D.~L.~Green, E.~F.~Jaeger, M.~A.~Jaworski, B.~P.~LeBlanc, A.~McLean, R.~Maingi, C.~K.~Phillips, L.~Roquemore, P.~M.~Ryan, S.~Sabbagh, G. 
Taylor, J.~R.~Wilson, Phys. Rev. Lett. {\bf 109}, 045001 (2012).

\bibitem{Perkins2} R.~J.~Perkins, J.~-W.~Ahn, R.~E.~Bell, A.~Diallo, S.~Gerhardt,
T.~K.~Gray, D.~L.~Green, E.~F.~Jaeger, J.~C.~Hosea, M.~A.~Jaworski, B.~P.~LeBlanc, G.~J.~Kramer, A.~McLean,
R.~Maingi, C.~K.~Phillips, M.~Podest`a, L.~Roquemore, P.~M.~Ryan, S.~Sabbagh, F.~Scotti, G.~Taylor, J.~R.~Wilson, Nucl. Fusion {\bf 53}, 083025 (2013). 

\bibitem{Bertelli} N.~Bertelli, E.~F.~Jaeger, J.~C.~Hosea, C.~K.~Phillips, L.~Berry, S.~P.~Gerhardt, D.~Green, B.~LeBlanc, R.~J.~Perkins,
P.~M.~Ryan, G.~Taylor, E.~J.~Valeo, J.~R.~Wilson, Nucl. Fusion {\bf 54}, 083004 (2014). 

\bibitem{Pace}D.~C.~Pace, R.~I.~Pinsker, W.~W.~Heidbrink, R.~K.~Fisher, M.~A.~Van~Zeeland, M.~E.~Austin, G.~R.~McKee, and
M. Garc$\acute{\text{i}}$a-Mu$\tilde{\text{n}}$oz. Nucl. Fusion {\bf 52}, 063019 (2012).

\bibitem{Wilson} J.~R.~Wilson, S.~Bernabei, T.~Biewer, S.~Diem, J.~Hosea, B.~LeBlanc, C.~K.~ 
Phillips, P.~Ryan, and D.~W.~Swain. Parametric Decay During HHFW on NSTX. AIP Conference 
Proceedings {\bf 787}, 66 (2005).

\bibitem{Porkolab1} M.~Porkolab, Nuclear Fusion 1978 {\bf 18},367 (1978).

\bibitem{Porkolab2} M.~Porkolab, Fusion Engineering and Design, {\bf 12}, 93 (1990).

\bibitem{Nieuwenhove} R.~V.~Nieuwenhove, G.~V.~Oost, J.~M.~Noterdaeme, M.~Brambilla, J.~Gernhardt,M.~Porkolab. Nucl.
Fusion {\bf 28}, 1603 (1988).

\bibitem{Fujii} T.~Fujii, M~Saigusa, H.~Kimura, M.~Ono, K.~Tobita, M.~Nemoto, Y.~Kusama, M.~Seki, S.~Moriyama, T.~Nishitani, H.~Nakamura, H.~Takeuchi, 
K.~Annoh, S.~Shinozaki, M.~Terakado and JT-60 team. Fusion Engineering and Design {\bf 12}, 139 (1990).

\bibitem{Rost} J.~C.~Rost, M.~Porkolab, R.~L.~Boivin, Phys. Plasmas {\bf 9}, 1262 (2002). 

\bibitem{Silin} V.~P.~Silin,  Zh. Eksp. Teor. Fiz. {\bf 48}, 1679 (1965); Sov. Phys. JETP {\bf 21}, 1127 (1965).

\bibitem{Porkolab} M.~Porkolab, Nuclear Fusion 1978 {\bf 18},367 (1978).

\bibitem{Mikhailenko1} V.~S.~Mikhailenko, K.~N.~Stepanov, Zh.~Eksp.~Teor.~Fiz. {\bf 87}, 161 
(1984)[Sov. Phys. JETP {\bf 60},  92 (1984).

\bibitem{Mikhailenko4} V.~S.~Mikhailenko, E.~E.~Scime, {\bf 11}, 3691 (2004).


\bibitem{Myra} J.~R.~Myra, D.~A.~D'Ippolito, D.~A.~Russell, L.~A.~Berry, E.~F.~Jaeger, M.~D.~Carter, Nuclear Fusion S455, {\bf 46} (2006). 





 

\bibitem{Mikhailenko2} V.~S.~Mikhailenko, V.~V.~Mikhailenko, K.~N.~Stepanov, Phys. Plasmas 
{\bf 18}, 062103 (2011).

\bibitem{Mikhailenko3} V.~V.~Mikhailenko, V.~S.~Mikhailenko, Hae~June~Lee, Phys. Plasmas 
{\bf 22}, 102308  (2015).

\bibitem{Dum} C.~T.~Dum, T.~H.~Dupree, Phys. Fluids {\bf 13}, 2064 (1971).
	
\bibitem{Benford} G.~Benford, J. Plasma Phys. {\bf 15}, 431 (1976).
\end{thebibliography}
\end{document}